\documentclass[aps,prb,twocolumn,superscriptaddress,longbibliography]{revtex4-2}

\usepackage{xcolor}	
\usepackage[utf8]{inputenc}
\usepackage{booktabs}
\usepackage{longtable}
\usepackage{siunitx}
\usepackage{amsmath}
\usepackage{mathrsfs}
\usepackage{graphicx}% Include figure files
\usepackage{dcolumn}% Align table columns on decimal point
\usepackage{bm}
\usepackage[normalem]{ulem}
\usepackage{float}
\usepackage{braket}% bold math
\usepackage{amssymb}
\usepackage{soul,xcolor}

\newcommand{\beq}{\begin{eqnarray}}
\newcommand{\eeq}{\end{eqnarray}}

\usepackage{hyperref}

\hypersetup{
colorlinks=true,
citecolor=blue,
linkcolor=blue,
urlcolor=blue,
}

\begin{document}

\title{EPR as a probe for the mechanisms of chirality-induced spin selectivity in photoinduced charge transfer}

\author{Arianna Cantarella}
\affiliation{Dipartimento di Scienze Matematiche, Fisiche e Informatiche, Universit\`a  di Parma, Parco Area delle Scienze, 7/A, I-43124 Parma, Italy.}
\affiliation{Gruppo Collegato di Parma, INFN-Sezione Milano-Bicocca, I-43124 Parma, Italy.}
\author{D. K. Andrea Phan Huu}
\affiliation{Dipartimento di Scienze Matematiche, Fisiche e Informatiche, Universit\`a  di Parma, Parco Area delle Scienze, 7/A, I-43124 Parma, Italy.}
\author{Paolo Santini}
\affiliation{Dipartimento di Scienze Matematiche, Fisiche e Informatiche, Universit\`a  di Parma, Parco Area delle Scienze, 7/A, I-43124 Parma, Italy.}
\affiliation{Gruppo Collegato di Parma, INFN-Sezione Milano-Bicocca, I-43124 Parma, Italy.}
\affiliation{UdR Parma, INSTM, I-43124 Parma, Italy.}
\author{Michael R. Wasielewski}
\email{m-wasielewski@northwestern.edu}
\affiliation{Department of Chemistry, Institute for Quantum Information Research
and Engineering, and Center for Molecular Quantum Transduction, Northwestern
University, Evanston, IL 60208-3113}
\author{Alessandro Chiesa}
\email{alessandro.chiesa@unipr.it}
\affiliation{Dipartimento di Scienze Matematiche, Fisiche e Informatiche, Universit\`a  di Parma, Parco Area delle Scienze, 7/A, I-43124 Parma, Italy.}
\affiliation{Gruppo Collegato di Parma, INFN-Sezione Milano-Bicocca, I-43124 Parma, Italy.}
\affiliation{UdR Parma, INSTM, I-43124 Parma, Italy.}
\author{Stefano Carretta}
\email{stefano.carretta@unipr.it}
\affiliation{Dipartimento di Scienze Matematiche, Fisiche e Informatiche, Universit\`a  di Parma, Parco Area delle Scienze, 7/A, I-43124 Parma, Italy.}
\affiliation{Gruppo Collegato di Parma, INFN-Sezione Milano-Bicocca, I-43124 Parma, Italy.}
\affiliation{UdR Parma, INSTM, I-43124 Parma, Italy.}

\begin{abstract} 
Several models for chirality-induced spin selectivity (CISS) in charge transfer have been proposed, but rarely compared directly with experiments. This step is essential for their validation and future progress of the field. 
%to validate different models starting from experimental observations. 
Here we investigate a broad class of radical pair states associated with different microscopic models by computing the related electron-paramagnetic resonance spectra. %Support to our analysis is provided by ab-initio calculations on prototypical molecules displaying CISS. 
This allows us to evidence peculiar signatures of different radical pair states, to study their experimental accessibility %, compare with experiments performed so far 
and identify conditions to verify proposed microscopic models.   
\end{abstract}

\maketitle

\section{Introduction}

The chirality-induced spin selectivity (CISS) effect encompasses a broad range of phenomena arising from the interplay between electron spin and molecular chirality. The first experimental manifestation of CISS was reported by Ray \textit{et al.}, who observed asymmetric scattering of photoelectrons transmitted through ordered monolayers of chiral organic molecules \cite{naaman_waldeck2012}.
Since then, CISS has been observed in a wide variety of chiral systems, ranging from individual molecules to molecular monolayers, multilayers, and crystals, and using several experimental techniques \cite{Naaman2019,Naaman2020}.

A large fraction of CISS experiments involves electron transport through a chiral material. 
%In a typical configuration, a chiral molecular layer, crystal, or single-molecule junction is placed between nonmagnetic and ferromagnetic contacts. The measured conductance then depends on the magnetization direction of the ferromagnetic component, indicating that charge transmission through the chiral material is spin selective. Reversing either the magnetization direction or the handedness of the molecular system generally reverses the observed asymmetry. Although these experiments have established CISS in a broad range of systems, the presence of electrodes and molecule--electrode interfaces complicates the identification of its microscopic origin. In particular, the measured response may depend not only on the intrinsic properties of the chiral molecule, but also on charge injection, interfacial electronic structure, electrode polarization, and nonequilibrium transport conditions.
More recently, signatures of CISS were also reported in photoinduced charge-transfer reactions \cite{Eckvahl2023}. In these experiments, a chiral organic bridge is covalently linked to electron-donor and electron-acceptor moieties. Following photoexcitation of the donor/acceptor, suitable energetic alignment among the donor (D), bridge, and acceptor (A) can drive charge transfer (CT) through the chiral bridge. For an initially closed-shell system, charge separation produces a pair of spatially separated radicals whose electron spins are correlated. The spin state of this radical pair can subsequently be investigated using time-resolved electron paramagnetic resonance (TREPR). For fast CT through an achiral bridge, the charge-separated radical pair state is a singlet, while introduction of a chiral  unit $\chi$ between D and A yields a triplet component which has therefore been used to estimate the contribution of CISS to radical-pair formation \cite{Eckvahl2023,Eckvahl2024,Latawiec2025,Latawiec2026}.

Photoinduced CT (PCT) offers several advantages over conventional transport experiments. First, the donor--bridge--acceptor architecture allows to investigate spin-selective charge transfer without metallic electrodes. The observation of CISS %-related spin polarization 
in such systems therefore demonstrates that metal contacts are not necessary for the effect, which arises %and places strong constraints on mechanisms that assign an essential role to molecule--metal interfaces. 
%These experiments are particularly striking because they involve 
in organic molecules characterised by a weak intrinsic spin--orbit coupling (SOC). This setup provides a stringent testing ground for theories seeking to explain the large spin selectivity attributed to CISS.

Moreover, preparation and detection of spin-polarized molecular states may create opportunities in quantum information science, for example by providing a chemically controlled route to electron-spin initialization and readout \cite{Aiello2022,Chiesa2023,Chiesa2025}. More broadly, understanding the interplay among chirality, charge transfer, and spin dynamics may be relevant to biological processes in which long-range charge transfer occurs through structurally chiral environments. Possible examples include biological energy conversion and radical-pair processes proposed to participate in avian magnetoreception \cite{hore2016radical,carmeli2016spin,Luo2021,naaman2022chiral,kobori2026vector}.

At the same time, the observation of CISS in PCT raises fundamental theoretical questions. Chief among these is how substantial spin polarization can emerge in organic systems characterized by weak SOC. Several microscopic mechanisms have been proposed, relying, in addition to SOC, on %spin--orbit-assisted CT, 
vibronic and dissipative effects \cite{Subotnik2022,Zhang2025}, many-body descriptions of the chiral bridge \cite{chiesa_many-body_2024}, effective donor-bridge spin--spin interactions \cite{Fay2021b,dmi_chiesa_2026,sarkar2026spinterface}, and the coupling of spin selectivity to molecular geometry \cite{zhou2026real}. In particular, a recently proposed model based on a vibrationally-mediated Dzyaloshinskii-Moriya interaction between donor and transferred electron spin predicts a triplet component in the radical pair state compatible with experimental observations \cite{dmi_chiesa_2026}.

Most theoretical studies have focused on the CT process itself, often in the absence of the external magnetic field used in TREPR. 
Their predictions are typically expressed in terms of a spin polarization or spin-dependent transfer probability, whereas the corresponding EPR spectrum is rarely calculated explicitly. Indeed, obtaining a sizeable spin polarization alone does not guarantee agreement with TREPR measurements, thus leaving an important gap between microscopic theories and experimental observables. %: the spectrum depends on the complete spin state prepared by electron transfer, including the direction of the polarization, the populations of the radical-pair spin levels, and any relevant coherences between them.

A complementary approach consists in 
%in has been adopted in a smaller number of studies in which the microscopic origin of CISS is not specified explicitly. Instead, 
treating CT as a state-preparation process that produces an assumed ``CISS state'', irrespective of its microscopic origin. The TREPR spectra are then simulated to determine the amount CISS contribution \cite{Chiesa2021,Luo2021,Ren2023,kobori2026vector}. Comparison with experiment can constrain which initial states %($\rho_{\mathrm{RP}}(0)$) 
are physically admissible. 
Such studies also identified experimental conditions that enhance the visibility of CISS and clarified how its signatures may survive orientational averaging. 
However, the initial states employed in these descriptions commonly rely on specific assumptions, including collinearity between the spin polarization, charge transfer, and radical pair dipolar axes and no net spin polarization, i.e. $\langle S_z\rangle=0$ ($S_z$ being the total spin projection along a chosen axes).  %in some geometries, equal populations of the (T$_+$) and (T$_-$) triplet sublevels}.
%This separation between microscopic state preparation and spectroscopic detection provides a useful framework for connecting theories of CISS to experiment. Irrespective of its microscopic origin, a CISS mechanism acting during electron transfer produces an initial radical-pair density matrix, ($\rho_{\mathrm{RP}}(0)$). Evolution of this state under the radical-pair spin Hamiltonian then generates the experimentally observed trEPR spectrum. 

In this work, we investigate the TREPR spectra arising from a broader class of initial radical-pair states compatible with spin-selective PCT and we connect the prepared state with microscopic mechanisms for CISS, to test them against experimental observations. We first relax (Sec. \ref{sec:nonColl}) the commonly adopted assumption that the spin polarization direction is collinear with the CT axis, examining its effects on both  orientationally ordered and disordered molecular ensembles. 
The examined state is related to exemplifying microscopic models, supported by electronic-structure calculations.
%We then use electronic-structure calculations and symmetry considerations to analyze the polarization directions predicted by exemplifying microscopic models. \\ 
%In particular, we consider whether these models can generate transverse polarization in chiral molecules and whether analogous contributions are symmetry allowed in achiral molecules belonging to low-symmetry point groups.
 We then investigate (Sec. \ref{sec:TpTm}) a prepared state with unequal populations of the $\ket{T_+}$ and $\ket{T_-}$ triplet sublevels, as could occur if the magnetic field applied along a generic direction in TREPR studies plays a role in the preparation of the radical-pair state \cite{dmi_chiesa_2026}.
%, in contrast to an assumption frequently adopted in existing trEPR descriptions of CISS. We therefore examine the spectroscopic consequences of field-dependent radical-pair preparation across different magnetic-field regimes.

By connecting microscopic models to the radical-pair states they produce, and these states to their TREPR signatures, we aim to establish which properties of CISS  can be constrained experimentally. 
%Rather than assessing a mechanism solely by the magnitude of the generated spin polarization, we ask whether it predicts the complete spin state required to reproduce the qualitative and quantitative features of the observed spectra. 
Finally, we discuss experimental conditions or molecular designs that could discriminate among  microscopic descriptions and clarify the origin of CISS in PCT.

\section{Non-collinear spin polarization and charge transfer}
\label{sec:nonColl}
\subsection{Computation of TREPR spectra}
Following photoexcitation and charge separation, a long-lived spin-correlated radical pair (SCRP), $D^{\bullet+}-B-A^{\bullet-}$, is formed. The two unpaired electron spins are predominantly localized on the donor and acceptor radical moieties, respectively, while the bridge mediates the preceding electron-transfer process. The spin dynamics of the charge-separated state is described by the radical-pair Hamiltonian
\begin{equation}\label{eq:scrp_hamiltonian}
    \hat{\mathcal{H}} = \frac{\mu_B}{h}\mathbf{B}_0 \cdot \mathbf{g}_D \cdot \hat{\mathbf{S}}_D + \frac{\mu_B}{h}\mathbf{B}_0 \cdot \mathbf{g}_A \cdot \hat{\mathbf{S}}_A + \hat{\mathbf{S}}_D \cdot \mathbf{d}_{DA} \cdot \hat{\mathbf{S}}_A
\end{equation}
where $\mathbf{B}_0$ is the external magnetic field, $\mathbf{g}_D$ and $\mathbf{g}_A$ are the $g$-tensors of the donor and acceptor radicals. The tensor $\mathbf{d}_{DA}$ describes the electron--electron interaction, including both isotropic and anisotropic (dipole-dipole) contributions. The effect of the hyperfine interaction with surrounding nuclear spins is here accounted for implicitly in the anisotropic $g$-tensors. 
%while $\hat{\mathcal{H}}_{HF}$ contains the hyperfine interactions between the electron spins and the magnetic nuclei. 
The dipolar axis approximately coincides with the vector connecting the centers of spin density of the donor and acceptor radicals. For well-localized and sufficiently separated radicals, this direction is also expected to provide a good approximation to the overall CT axis. 
%The two directions will therefore be identified, unless stated otherwise. da qualche parte facciamo diversamente? Se no toglierei specifica

The second essential ingredient is the density matrix describing 
the spin state prepared upon formation of the radical pair, $\rho_{\mathrm{RP}}(0)$, 
that encodes the outcome of the preceding CT process. 
In this sense, the microscopic mechanism of CISS acts as a state-preparation mechanism: different microscopic descriptions may produce radical pairs with different spin states, both in terms of populations and coherences between singlet and triplet states. The TREPR spectrum is then computed by rotating the prepared state  $\rho_{\mathrm{RP}}(0)$ 
in the basis of the Hamiltonian \eqref{eq:scrp_hamiltonian} eigenstates and computing transition intensities between the eigenstates in linear response theory (after complete coherence loss). These are affected by the initial state via their population difference. Since we focus on fast (sub-nanosecond) CT processes \cite{Fay2021}, it is perfectly  reasonable to assume a pure spin state $\rho_{\mathrm{RP}}(0)$ and then neglect coherences after projection of this state on the radical-pair eigenbasis.

Experimental analyses often neglect the microscopic origin of $\rho_{\mathrm{RP}}(0)$. A restricted, parametrized form of $\rho_{\mathrm{RP}}(0)$ is assumed, and its parameters are determined by comparison with the measured spectrum, thus constraining the family of admissible radical pair states.
%Previous works on detection of CISS using EPR have generally represented its effect through a restricted parametrization of the initial radical pair state.  
A commonly employed description to simulate the TREPR spectra of CISS-polarized radical pairs was introduced by Hore \textit{et al.} and subsequently generalized \cite{Fay2021,Luo2021,Ren2023,Eckvahl2023,Eckvahl2024,Latawiec2025,Chiesa2025}. In this case, the radical pair is prepared in a coherent superposition of the singlet $\ket{S}$ and triplet $\ket{T_0}$ states,
\begin{equation}
\label{eq:hore-state}
\left|\psi_{\parallel}^\chi \right\rangle
=
\cos\left(\frac{\chi}{2}\right)\left|S\right\rangle
+
e^{i \gamma} \sin\left(\frac{\chi}{2}\right)\left|T_0\right\rangle ,
\end{equation}
where (in the product basis) $\ket{S}=(\ket{\uparrow\downarrow}-\ket{\downarrow\uparrow})/\sqrt{2}$ and $\ket{T_0}=(\ket{\uparrow\downarrow}+\ket{\downarrow\uparrow})/\sqrt{2}$ for a  quantization direction parallel to the CT and dipole-dipole axes in Eq. \eqref{eq:scrp_hamiltonian}.    
%
%\begin{equation}
%\left|S\right\rangle
%=
%\frac{1}{\sqrt{2}}
%\left(
%\left|\alpha_D\beta_A\right\rangle
%-
%\left|\beta_D\alpha_A\right\rangle
%\right)
%\end{equation}
%
%and
%
%\begin{equation}
%\left|T_0\right\rangle
%=
%\frac{1}{\sqrt{2}}
%\left(
%\left|\alpha_D\beta_A\right\rangle
%+
%\left|\beta_D\alpha_A\right\rangle
%\right).
%\end{equation}
%
%Here, $\alpha$ and $\beta$ denote the electron-spin states parallel and antiparallel, respectively, to the selected quantization axis. 
The parameter $\chi$ controls the amount and sign of CISS %degree and sign of the spin polarization
generated during charge separation. Spin-conserving electron transfer corresponds to $\chi=0$. At the limiting values $\chi=\pi/2$ and $\chi=-\pi/2$, the radical pair is prepared in the fully spin-polarized product states $\ket{\uparrow\downarrow}$ and $\ket{\downarrow\uparrow}$, respectively. 
Intermediate values of $\chi$ interpolate continuously between the singlet-born and fully spin-polarized limits. The CISS efficiency assessed by TREPR is related to the triplet population by $p=(1-\cos \chi)$. Note that EPR experiments performed on isotropic solutions or samples aligned with liquid crystals (where both orientations of the molecule are present for any given direction) are not sensitive to the relative phase between singlet and triplet and hence the parameter $\gamma$ is still undetermined. 

The state \eqref{eq:hore-state} is restricted to the $\langle S_z\rangle =0$ subspace and, as denoted by the subscript  $\parallel$, the CISS axis is assumed to coincide with the dipolar and CT axes. The molecular orientation relative to the external magnetic field may vary (and the state $\ket{\psi^\chi_\parallel}$ is rotated accordingly), but the polarization, dipolar, and charge-transfer axes remain mutually collinear.

The spectral consequences of this initial state have been investigated for both oriented and isotropic ensembles of radical pairs.\cite{Chiesa2025}
For typical systems, a CISS state $\ket{\psi^\chi_\parallel}$ gives reversed spectral features compared to the singlet when the magnetic field is applied perpendicular to the dipolar axis, while the two states yield practically equivalent spectra when the samples are aligned with liquid crystals parallel (and anti-parallel) to the field. 
This, in turn, can give clear signatures of CISS even in a randomly oriented solution if spectral features originating from differently oriented molecules are spectroscopically resolved and/or the perpendicular contributions dominate the spherical average \cite{Chiesa2021,Ren2023}. 

%In the absence of a CISS contribution, a singlet-born radical pair produces the characteristic multiplet polarization of a spin-correlated radical pair, generally manifested as an antiphase pattern of emissive and absorptive transitions. 
% The CISS contribution changes the relative populations of the radical-pair spin states and introduces a net polarization component, thereby modifying the relative intensities and, in some cases, the signs of the EPR transitions.

%For an isotropic ensemble, one might expect the CISS contributions from oppositely oriented molecules to cancel. However, in the presence of an anisotropic dipolar interaction, oppositely polarized molecular orientations do not necessarily contribute at the same positions in the spectrum. Consequently, the cancellation is generally incomplete, and CISS-induced changes can survive orientational averaging. Previous simulations have shown that these contributions are particularly visible when at least one radical has a small $g$-anisotropy and an inhomogeneous linewidth larger than the dipolar splitting. Oriented samples are nevertheless expected to provide more direct access to the magnitude and angular dependence of the CISS polarization.
%\textbf{Dire brevemente cosa si vede con questo stato facendo riferimento alla letteratura }
Equation~\eqref{eq:hore-state} provides a phenomenological description of CISS-polarized radical pairs, but it also introduces a specific relation between the spatial direction of charge transfer and the prepared spin state. In particular, the assumption of a spin polarization parallel to the dipolar/CT axis is not a necessary consequence of molecular chirality.
%\textbf{Dire che ci sono vari modelli microscopici che pero' sembrano puntare al fatto che la polarizzazione possa non essere necessariamente nella direzione del CT/dipolare.}
As discussed below, microscopic mechanisms in which the polarization direction is controlled by SOC (or by a related effective spin--spin interaction) may predict a polarization axis that is not parallel to the CT direction. This motivates extending the conventional initial state description to radical pairs in which the polarization and dipolar axes are independently oriented. Therefore, we perform the following TREPR simulations using an initial coherent state of the form
\begin{equation}\label{eq:rotated_initial_state}
    \ket{\psi^\chi_{\alpha,\beta}} = R(\alpha,\beta) \ket{\psi^\chi_\parallel} \equiv R_z(\beta)R_y(\alpha) \ket{\psi_\parallel^\chi} ,
\end{equation}
which corresponds to rotating the spin polarization from the dipolar axis $z$ to a generic orientation parametrized by polar angles $(\alpha,\beta)$.

\subsection{Simulations}
To investigate how a broad class of initial radical pair states influences TREPR spectra 
without focusing on a specific system (with subtle anisotropies and parameters strain), we consider a simple prototypical molecule with axial ${\bf g}$ tensors typical of radicals employed in this context. %, such as PXX$^+$ or BDX$^+$ donors and NDI$^-$ acceptor. Check o togliere nomi se troppo specifici. 
We fix ${\bf g}_A=(2.0045, 2.0045, 2.0025)$ and ${\bf g}_D=(2.0032, 2.0032, 2.0021)$ and we include an anisotropic dipole-dipole interaction computed in the point dipole approximation for two spins at a distance of 2 nm. %We then compare with specific experimental systems 
The considered $g$-anisotropy is perfectly realistic \cite{Eckvahl2023,Eckvahl2024,Latawiec2025,Latawiec2026}  and sufficient to resolve in Q band spectral features from different orientations even in isotropic solution. 

%we adopt the donor-bridge-acceptor systems reported in Ref.~\citenum{Eckvahl2024}. These systems offer several features relevant for the present analysis: both chiral and achiral variants were presented and designed to be sufficiently similar in terms of EPR parameters, and their spin Hamiltonian is well characterized, with sufficient $g$-anisotropy to resolve distinct spectral features across the magnetic field even in isotropic solution.\andreap{la parte su H anisotropo solo per SOC la menzioniamo dopo, perche' e' vera per tutti i sistemi di Mike, e ha piu' che a che fare con il modello microscopico che con l'SCRP} 

%In this work, \andreap{unless otherwise stated,} we adopt the same parameters for the SCRP Hamiltonian \eqref{eq:scrp_hamiltonian} as in Ref.~\citenum{Eckvahl2024} (Table XXX), \andreap{removing for simplicity the coupling with nuclear spins in X band and correspondingly increasing the peak-to-peak linewidth to 0.2 mT}. 
 %direction of SOC indicates also the direction of spin polarization according to different models (only anisotropic term in H apart from B).} \\

\begin{figure*}[ht]
    \centering
    \includegraphics[width=\linewidth]{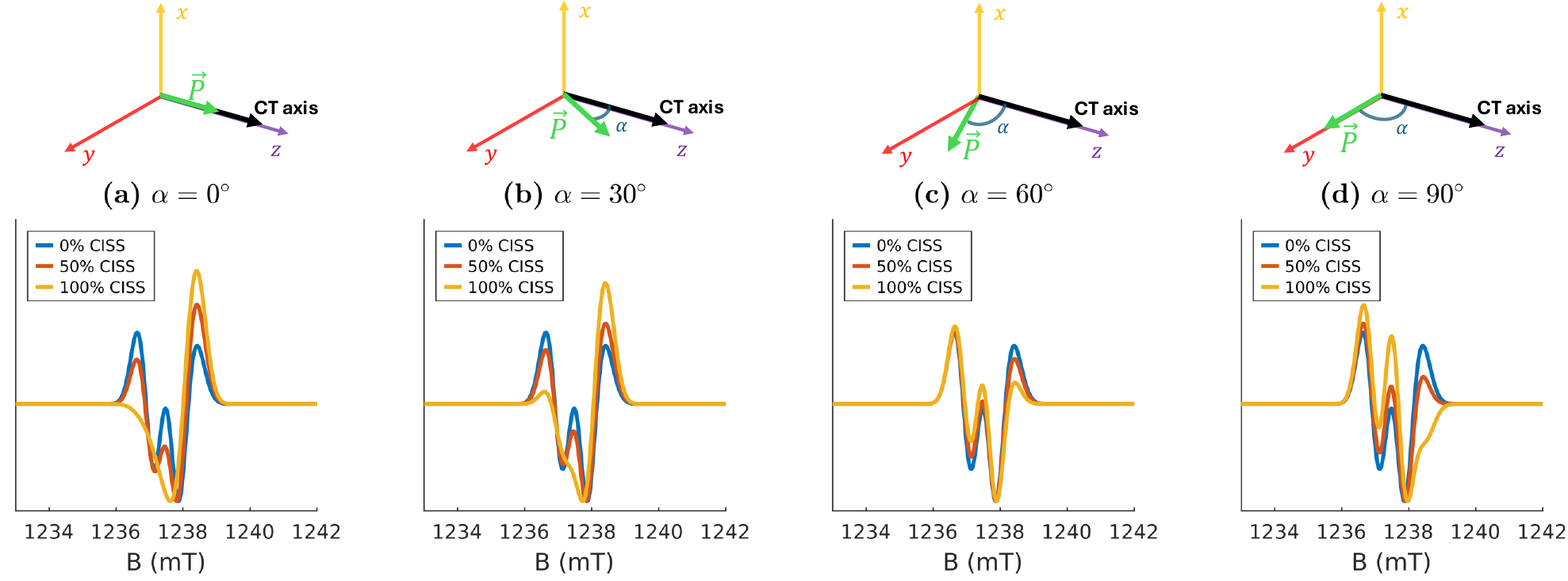}
    \caption{\textbf{Simulated powder TREPR spectra for an initial triplet polarization along different molecular-frame directions for a prototypical molecule in Q-band.} The directions are parametrized by the polar angle $\alpha$ from the CT axis ($\beta = 0$ throughout). Panel (a): polarization direction parallel to the CT axis ($\alpha=0^\circ$); panel (d): polarization direction orthogonal to the CT axis ($\alpha=90^\circ$); panels (b,c) show intermediate orientations. Each panel reports spectra for CISS contributions of 0\%, 50\%, and 100\%.}
    \label{fig:toy_powder_direction}
\end{figure*}

We begin by examining how the relative orientation of the dipolar axis and the polarization direction affects the EPR spectrum of both isotropic and oriented samples. Fig.~\ref{fig:toy_powder_direction} 
shows the simulated TREPR spectra of randomly oriented samples, computed using Eq.~\eqref{eq:rotated_initial_state} as the initial state for different values of $\alpha$ (with $\beta=0$) and different CISS efficiencies. Fig.~\ref{fig:toy_powder_direction}a represents the case in which the dipolar axis and the polarization direction are collinear, as in Eq.~\eqref{eq:hore-state}. 
%and as assumed in the trEPR simulations of Ref.~\citenum{Eckvahl2024}; 
Here the CISS-driven features at low field (yellow curve) are clearly distinguishable from the singlet signal (in blue). A similar analysis   holds for Fig.~\ref{fig:toy_powder_direction}b, where the polarization axis is  tilted relative to the dipolar axis by $\alpha=30^{\circ}$. Also in this case, the CISS features remain distinguishable from the singlet. By contrast, in Fig.~\ref{fig:toy_powder_direction}c–d the CISS-driven and singlet spectra nearly overlap. However, we note an inverted peak at high-field for $\alpha=90^\circ$.
These results indicate that in isotropic solution
the main spectral features come from the polarization component parallel to the dipolar axis. A triplet component orthogonal to the dipolar axis could be present without yielding clear spectral features (unless with high resolution). 
%Per nostro claim forse piu chiare simulazioni banda X, che dipendono meno dai dettagli dei g, check. Non è che singoletto e CISS siano simili, devi guardare sempre 100\%. Al 50 sono simili solo perche singoletto pesa un po piu del ciss, ma quello è un dettaglio dovuto ai parametri. Basta che rifai con 60-70\% CISS o con g piu simili.

\begin{figure*}[ht!]
	\centering
\includegraphics[width=0.7\linewidth]{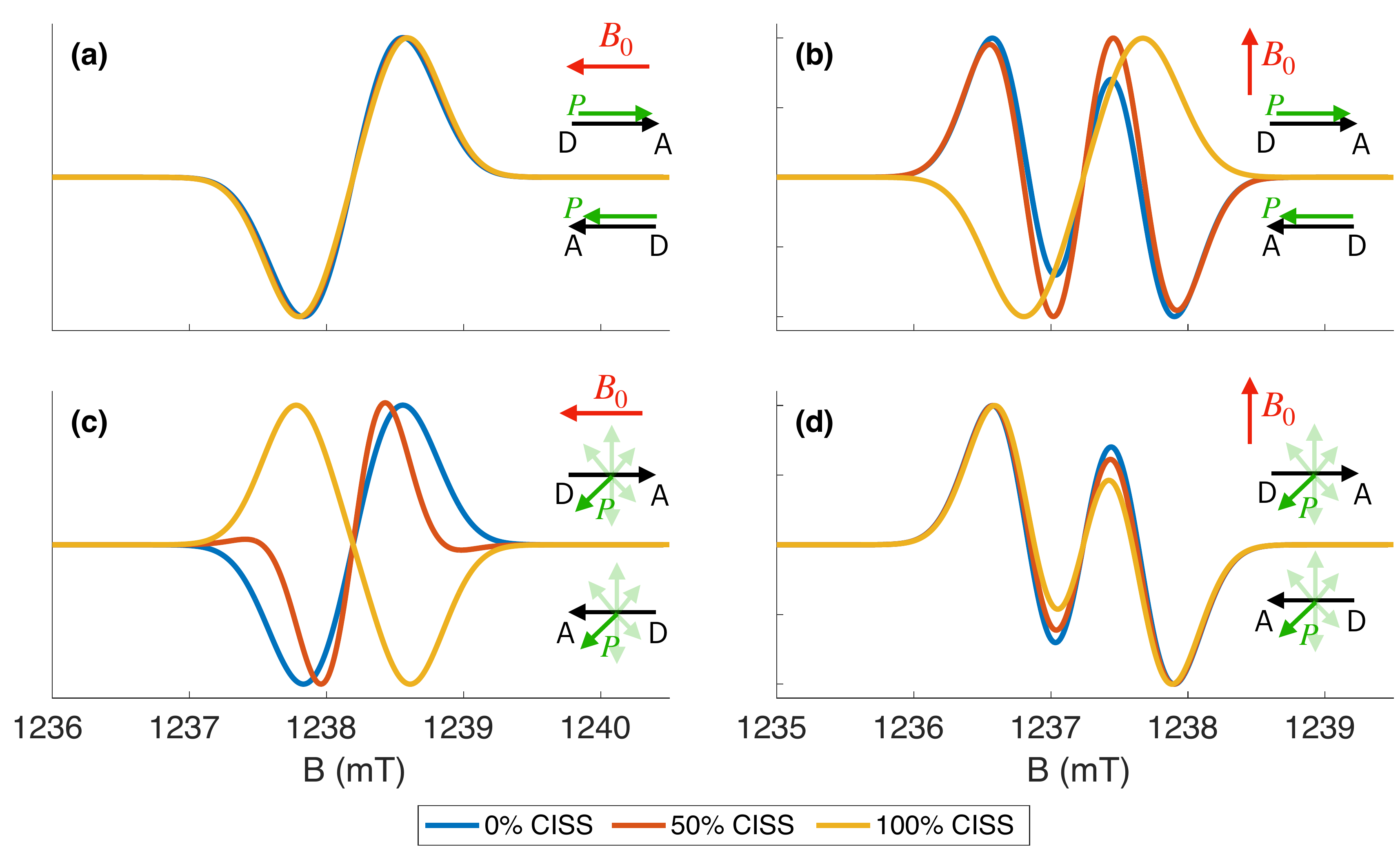}
	\caption{\textbf{Simulated oriented-sample TREPR spectra for the four combinations of polarization orientation relative to the dipolar axis, for a prototypical molecule, in Q-band.} Rows: orientation of the polarization vector relative to the molecular dipolar axis (parallel, top; perpendicular, bottom). Columns: orientation of the external magnetic field $\mathbf{B}_0$ relative to the dipolar axis ($\mathbf{B}_0 \parallel$, left; $\mathbf{B}_0 \perp$, right). In each panel, spectra are shown for CISS contributions of 0\%, 50\%, and 100\%. }
	\label{fig:toy_epr_oriented}
\end{figure*}

A deeper insight into the effect of a polarization noncollinear with the dipolar axis can be gained by considering experiments on samples aligned with liquid crystals (LC). To reproduce these experimental conditions, we include in each simulation sets of oppositely oriented molecules. 
%, i.e. $\phi=0,\pi/2,\pi,3\pi/2$ for $\theta=\pi/2$ and $\theta=0,\pi$ in the bottom lines.
The corresponding simulations are shown in Fig.~\ref{fig:toy_epr_oriented}. 
We considered four configurations, defined by the relative orientation of the magnetic field with respect to the dipolar axis, and of the dipolar axis with respect to the polarization direction.
Fig.~\ref{fig:toy_epr_oriented}a shows the simulated spectra for the case in which all three directions are collinear; no evident CISS features emerge. In contrast, Fig.~\ref{fig:toy_epr_oriented}b corresponds to a magnetic field orthogonal to the dipolar axis, with the dipolar axis parallel to the polarization direction: here, a clear inversion of the signal is observed for 100\% CISS contribution, along with different intensities for the 50\% CISS case. This is the configuration in which CISS was evidenced for the first time \cite{Eckvahl2023} and provides the most important contribution to the spectral signatures displayed in Fig.~\ref{fig:toy_powder_direction}a.
We now consider the complementary case in which the polarization direction is perpendicular to the dipolar axis, i.e. we fix $\alpha=\pi/2$ and we average over $\beta$ in Eq. \eqref{eq:rotated_initial_state}, to account for the polarization direction which remains unconstrained in the plane perpendicular to the dipolar axis when the solvent is frozen. 
%Experimentally, molecules aligned in a frozen solvent are constrained only along their dipolar axis, while the two orthogonal directions remain unconstrained. Consequently, when simulating a polarization direction orthogonal to the dipolar axis, one must account for the fact that this direction is not fixed within the plane perpendicular to the dipolar axis, but is instead free to rotate about it. To do so, we averaged over the trEPR signal obtained from preparing the initial state as in Eq.~\eqref{eq:rotated_initial_state}, with $\alpha=90^{\circ}$ and $\beta \in [0, 2\pi)$. 
Results are shown in Fig.~\ref{fig:toy_epr_oriented}c-d for the magnetic field parallel and perpendicular to the dipolar axis, respectively. 
Notably, for a magnetic field parallel to the dipolar axis and orthogonal to the polarization direction (Fig.~\ref{fig:toy_epr_oriented}c) we observe clear CISS features, differing from the singlet.  %Moreover, this results is opposite to what observed in an isotropic solution, for which the case with $\alpha=90^{\circ}$ was not showing any notable difference between CISS and no CISS spectra. 
Finally, no evident CISS features are observed for a magnetic field orthogonal to the dipolar axis and to the polarization reported in 
Fig.~\ref{fig:toy_epr_oriented}d. The two configurations of Fig.~\ref{fig:toy_epr_oriented}c and Fig.~\ref{fig:toy_epr_oriented}d are directly comparable to Fig.~\ref{fig:toy_powder_direction}d. When $\alpha=90^\circ$ we can observe CISS signature for samples oriented along the field, but in the spherical average configurations with the dipolar axis orthogonal to the field dominate and make the visibility of the effect much weaker.

In summary, while measurements on an isotropic solution mainly detect the component of the polarization parallel to the dipolar axis, chiral molecules oriented with LC can provide a CISS signal distinguishable from the singlet when the magnetic field is orthogonal to the polarization vector. 
%under two conditions: (i) when the magnetic field is perpendicular to the dipolar axis and a significant polarization is along the same dipolar axis; or (ii) when the magnetic field is parallel to the dipolar axis, with a significant polarization orthogonal to it. 
%The converse problem (inferring the polarization direction from the EPR features) is less straightforward, as it also depends on the magnitude of the spin polarization itself.

\begin{figure}
	\centering
	\includegraphics[width=.85\linewidth]{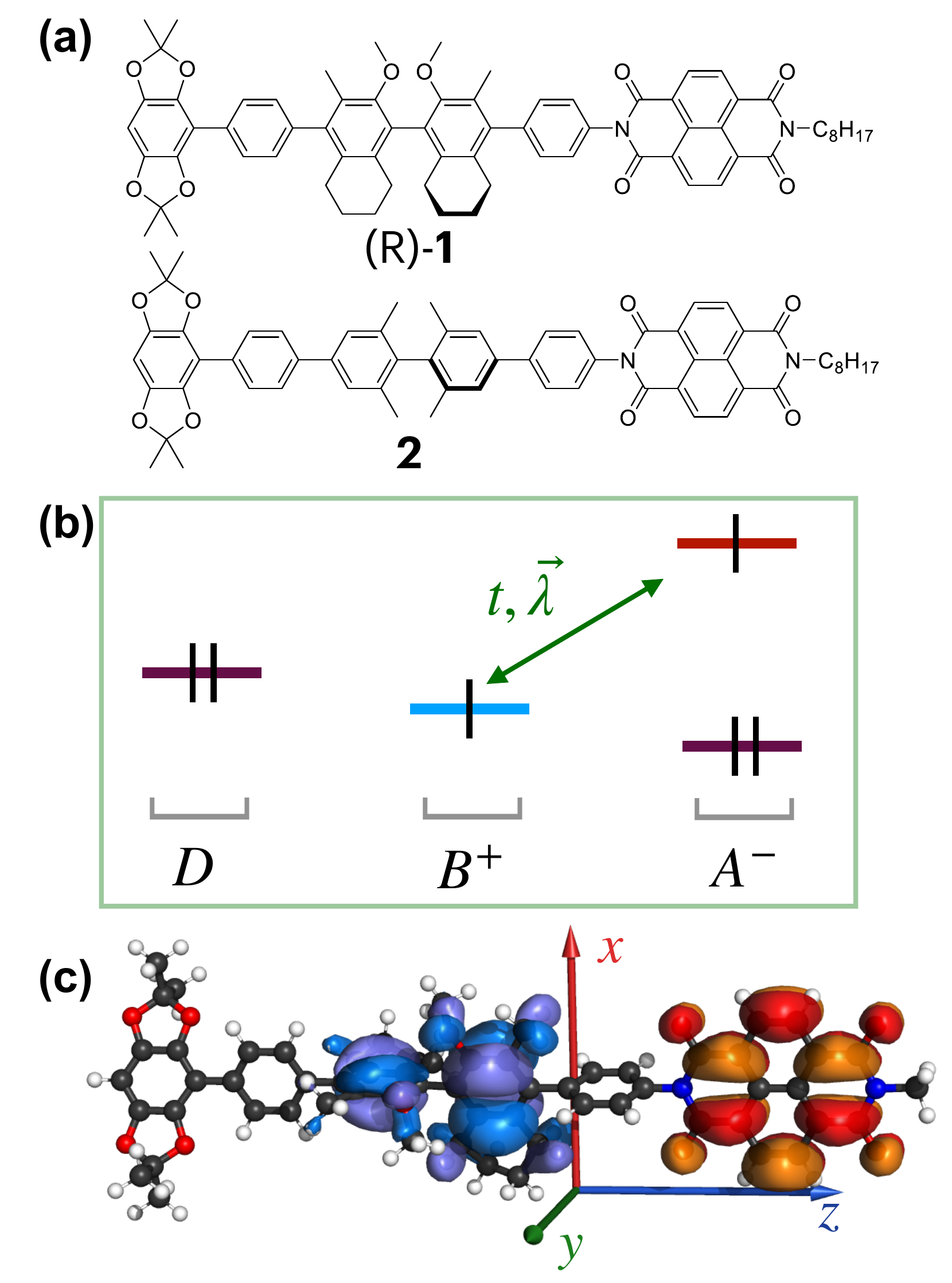}
	\caption{
    % \textbf{Energy-level scheme and frontier orbitals of the donor–bridge–acceptor (D–B\textsuperscript{+}–A\textsuperscript{-}) model system.} 
    \textbf{(a):} molecular structures of the chiral and achiral donor–bridge–acceptor systems. (\textit{R})-\textbf{1}: (\textit{R}) enantiomer of the chiral molecule presented in Ref.~\cite{Eckvahl2024}. \textbf{2}: achiral control compound from Ref.~\cite{Eckvahl2024}.
    \textbf{(b)} scheme of MOs energy levels, representing the D–B\textsuperscript{+}–A\textsuperscript{-} configuration, with the orbital localized on the bridge (blue) and the one one the acceptor (red) coupled via the hopping matrix element $t$ and the spin-orbit coupling vector $\vec{\lambda} = (\lambda_x, \lambda_y, \lambda_z)$, with values $t=0.16$ meV and $\vec{\lambda} = [0.0293\,\hat{x}, -0.0003\, \hat{y}, -0.0060\, \hat{z}]\,$ meV. \textbf{(c):} molecular structure of the D-B-A system, with isosurfaces of the relevant frontier orbitals overlaid: the bridge orbital (blue) and the acceptor orbital (orange/red). The isovalue of the isosurfaces is 0.02 $\AA^{-3}$.}
	\label{fig:scheme_coupling_jacs}
\end{figure}

We have checked that the same conclusions drawn here for typical Hamiltonian parameters and experimental conditions %(bands, linewidth)
also hold for a larger class of systems. In particular, we have examined the BDX-B$_\chi$-NDI molecules reported in Ref.~\citenum{Eckvahl2024} and shown in Fig.~\ref{fig:scheme_coupling_jacs}a. These systems offer several properties relevant for the present analysis: both chiral and achiral variants were synthesized and designed to be very similar in terms of EPR parameters, and their spin Hamiltonian is well characterized. The TREPR simulations in X-band for the isotropic solution and the oriented sample are reported in Fig.~\ref{fig:jacs_powder_direction} and Fig.~\ref{fig:jacs_epr_oriented} of the Supplementary Information, respectively.
Experimental measurements were only reported on randomly oriented samples, hence they cannot evidence a CISS polarization component orthogonal to the dipolar axis.\cite{Eckvahl2024} As discussed above, proving this would require experiments on samples aligned with LC. \\

% \begin{figure}
%     \centering
%     \includegraphics[width=.85\linewidth]{molecules_kekule}
%     \caption{\textbf{Molecular structures of the chiral and achiral donor–bridge–acceptor systems.} (\textit{R})-\textbf{1}: (\textit{R}) enantiomer of the chiral molecule presented in Ref.~\cite{Eckvahl2024}. \textbf{2}: achiral control compound from Ref.~\cite{Eckvahl2024}.}
% \label{fig:molecular_structures}
% \end{figure}
%Moreover, these results highlight one of the key advantages of measurements on oriented samples: in a powder spectrum, only a polarization component along the dipolar axis is detectable, since a polarization orthogonal to it, even if present, within this system should be washed out by the spherical average, as shown in Fig.~\ref{fig:powder_direction}d. Oriented-sample measurements, by contrast, can reveal such a polarization component, provided the dipolar axis is aligned parallel to the magnetic field.
%Si puo mettere anche hairpins, unico difetto e' mancanza achirale. Ma forse li metterei. Science non lo menzioniamo ma svantaggi sono che features in direzioni diverse tendono a sovrapporsi, spettro dipende criticamente da iperfini non ben caratterizzati \\
%Dopo diciamo che nell'achirale rispetto al lavoro \cite{dmi_chiesa_2026} il SOC e hopping non sono mai tra stessi orb e quindi darebbero DM nulla (statica) ma esplorare simmetrie vibrazioni va oltre questo lavoro. In ogni caso direzione del SOC nel chirale e' generica .

\subsection{Comparison with microscopic models via \textit{ab initio} calculations}
Several proposed microscopic models for CISS in CT allow for polarization directions non-collinear with the dipolar axis. 
Indeed, in most models (in zero or small magnetic field) the polarization direction (or more generally the quantization axis of the triplet component) is determined by the SOC, which is the only anisotropic term in the Hamiltonian. 
This is also the case of models based on spin--spin interactions between the donor and transferred spin, such as those proposed in  Refs.~\citenum{Fay2021b} and \citenum{dmi_chiesa_2026}. 
%This is also true for models that do not rely on SOC, like the spinterface model, which has been recently proposed also in the context of CISS in PET.\cite{sarkar2026spinterface} 
%There, the polarization direction is completely defined by the direction of an effective solenoid field, that only approximately can be considered to be collinear with the CT axis. 
In the following discussion we consider these models as representative cases to analyse a spin polarization potentially non-collinear with the CT axis. 

We have recently shown \cite{dmi_chiesa_2026} that the interaction between the transferred electron and the one on the donor can be recast into an effective spin Hamiltonian including isotropic, anisotropic and Dzyialoshinskii-Moriya spin-spin interactions. These couplings are effective in multi-step CT processes (typical of molecules where CISS was observed so far \cite{Eckvahl2023,Eckvahl2024,Latawiec2025,Latawiec2026}), when the transferred electron spends a sufficient amount of time on the chiral bridge. Notably, the model explains with realistic parameters the observed triplet spin component in the radical pair. 
The origin of these spin-spin interactions are virtual excitations via spin independent (hopping) and spin-dependent (SOC) processes and its most important contribution is mediated by molecular vibrations \cite{dmi_chiesa_2026}.
Nonetheless, a static (not vibrationally mediated) contribution leading to the same spin Hamiltonian was first introduced in \cite{Fay2021b}. %, who showed that the interplay between donor-bridge hopping, exchange interaction and SOC under incoherent multi-step CT can generate a spin polarization. 
In both Refs.~\cite{Fay2021b,dmi_chiesa_2026} the polarization direction is dictated then by the Dzyialoshinskii-Moriya interaction (DMI) vector, whose orientation in turn depends on the direction of the SOC pseudovector and of its modulation by molecular vibrations. 
Although the vibrationally-mediated contribution should dominate, given  %considering 
the cost of computing the modulation of SOC due to many vibrational modes, we consider here the static contribution to SOC as a first case study to address its possible orientation with respect to the dipolar axis. 

%qualitative considerations can be made by only addressing the purely electronic terms, vastly simplifying the following first principle calculations. 

As a concrete example, we perform first-principles calculations on the BDX-B-NDI molecules reported in Ref.~\citenum{Eckvahl2024}. The system \textit{R}-\textbf{1} and its achiral reference \textbf{2} are shown in Fig.~\ref{fig:scheme_coupling_jacs}a. Since (\textit{R})-\textbf{1} and \textbf{2} are known to undergo photoinduced \textit{hole} transfer, we evaluate by DFT the SOC and hopping matrix elements coupling a bridge-localized occupied frontier orbital to an acceptor-localized unoccupied frontier orbital (Fig.~\ref{fig:scheme_coupling_jacs}b-c). Details of the calculation are given in Appendix~\ref{app:localization}.
The resulting SOC pseudovector for this orbital pair is $\vec{\lambda} = [0.0293\,\hat{x}, -0.0003\,\hat{y}, -0.0060\,\hat{z}]$~meV: predominantly transverse to the CT direction, with a small but non-negligible longitudinal component.

The $D$--$B^{\bullet+}$--$A^{\bullet-}$ intermediate state of the photoinduced hole transfer process is subject to a DMI proportional to $t \vec{\lambda}$~\cite{dmi_chiesa_2026}, and hence oriented along the same direction as the SOC pseudovector computed above. This result suggests a preferred spin-polarization direction that is not collinear with the CT direction.However, we remark that the present calculation captures only the purely electronic contribution to the effective DMI; vibronic pathways can add further contributions, potentially altering both the magnitude and direction of the total DMI vector. 
%Experimental results on isotropic solutions~\cite{Eckvahl2024}, mostly sensitive to the polarization component parallel to CT, suggest that vibrational contributions could be leading, motivating further study in this direction.
%per quanto, se polvere e' sensibile solo al parallelo vuol dire che parallelo ci deve essere (piu grosso di quello che calcoli), ma non so se deve essere leading.

For the achiral, $C_{2v}$ reference \textbf{2},
symmetry forbids the electronic DMI contribution, consistently with the original Moriya's analysis \cite{Moriya}. Indeed, no pair of orbitals can be simultaneously coupled via hopping and SOC, as shown in the Supplementary Fig.~\ref{fig:c2v_matrices}. 

An interesting study which could lead to a polarization tilted from the CT axis was reported in \cite{Agnoloni2026}, where the Authors synthesized a molecule including a helicene donor and undergoing photoinduced hole transfer. Measurements are compatible with a small CISS efficiency with quantization along the helicene axis (about $60^\circ$ from the dipolar axis). However, it is worth noting that in that case the chirality is only involved in the last CT step. Hence, it likely does not allow for a significant triplet component to arise in the intermediate CT step as proposed by models  \cite{dmi_chiesa_2026,Fay2021b}.

\begin{figure*}[ht!]
	\centering
	\includegraphics[width=0.85\linewidth]{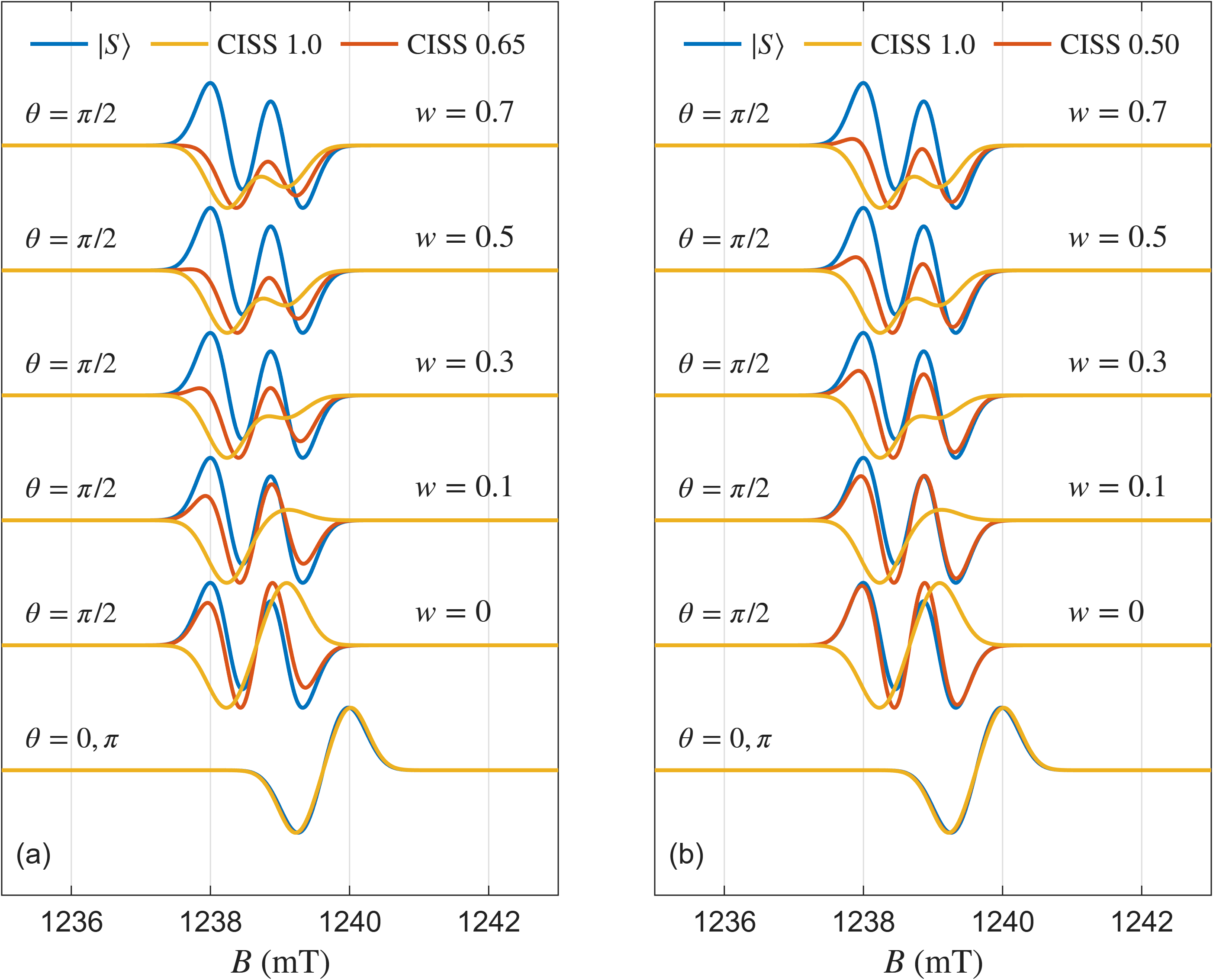}
	\caption{Simulation of TREPR spectra on the benchmark donor-bridge-acceptor molecule with spin Hamiltonian parameters ${\bf g}_A=(2.0045, 2.0045, 2.0025)$, ${\bf g}_D=(2.0032, 2.0032, 2.0021)$ and anisotropic dipole-dipole interaction ${\bf D}_{DA}=(1,1,-2)\times 6.5$ MHz, corresponding to a DA distance of 2 nm. In each panel, different set of curves refer to different values of $\ket{T_+}-\ket{T_-}$ population imbalance $w$ for $\theta=\pi/2$ (including oppositely oriented molecules in LC alignment). The bottom lines are a reference for $\theta=0$, not depending on $w$. Black lines are computed starting from a singlet radical pair state, blue curves for perfect (100\%) CISS efficiency, red ones for 65 or 50\% CISS efficiency (a vs b panel). Each spectrum is normalized to the minimum. Simulations are performed using Matlab code easyspin-6.0.0-dev.53.}
	\label{fig:TpTmtoy}
\end{figure*}
\section{Observing a net spin polarization}
\label{sec:TpTm}
Another interesting point is the spectroscopic effect of a net spin polarization
(i.e. a different population of the $\ket{T_+^L}$ and $\ket{T_-^L}$ high-field sublevels) when the magnetic field is perpendicular to the dipolar axis. We remind that this configuration is the most susceptible to CISS effect in LC measurements \cite{Eckvahl2023,Chiesa2025}.
When a significant magnetic field is applied, it is convenient for the interpretation of EPR experiments to work in the {\it laboratory} frame, where the magnetic field is along $z_L$ and the molecules are in a generic orientation. We refer hereafter to the triplet sublevels rotated in the laboratory frame with the superscript $L$.
This population imbalance between  $\ket{T_+^L}$ and $\ket{T_-^L}$ could arise according to some of the proposed CISS models. For instance, 
%a state $\ket{S}+\ket{T_0}+\ket{T_+}$ 
it was postulated in Ref. \cite{Matxain2019} (for both transport and ET  through a chiral molecule) and can also emerge in presence of a sizeable external magnetic field in the model proposed in Ref. \cite{dmi_chiesa_2026}, and already introduced in Sec. II.B. In the following discussion we parametrize the isotropic and Dzyialoshinskii-Moriya interactions by $J$ and $D$  and we neglect for simplicity the typically smaller anisotropic $J_a$ contribution. We recall that these interactions occur in the intermediate CT step (when the transferred electron is still on the bridge) and are negligible in the final radical-pair Hamiltonian \eqref{eq:scrp_hamiltonian}. 
%As outlined above, these interactions originate from a  combination of (static and vibrationally modulated) hopping and SOC between donor HOMO and bridge LUMO orbitals (or acceptor LUMO and bridge HOMO in the case of hole transfer). 
As already discussed, the direction of the DMI vector driving the mixing between singlet and triplet observed by EPR  is given by the SOC. To simplify the following discussion, hereafter we assume it parallel to the CT axis. 

In zero field, the DMI brings the initial singlet state into a superposition of singlet and triplet, where the quantization axis is determined by the DMI pseudovector. The triplet population at the end of the CT depends on the ratio between $D$ (which mixes singlet and triplet) and $J$ (which splits them) 
and on the CT time. The latter determines when  the singlet-triplet coherent oscillation is stopped to accumulate charge on the acceptor. \\
In an applied field and for generic orientation of the molecules parametrized by polar angles $\{\theta,\phi\}$ in the laboratory frame,  the DMI vector includes all the three components ${\bf D}=(D \sin \theta \cos \phi, D \sin \theta \sin \phi, D \cos \theta)$ and hence the initial photoexcited singlet state can be correspondingly mixed with $\ket{T_0^L}$, $\ket{T_+^L}$ and $\ket{T_-^L}$, with coupling constants $D_z$ and $D_y \pm iD_x$, respectively. Note that at fixed CT time the mixing of the singlet with each of the triplet sublevels will also depend on their different energy gaps. In particular, $\ket{S}$ and $\ket{T_0^L}$ are still separated by an energy gap $J$, while $\ket{T_+^L}$ and $\ket{T_-^L}$ differ by twice the Zeeman energy. For ferromagnetic isotropic exchange as in Ref. \cite{dmi_chiesa_2026}, triplet states are energetically favored in zero field and hence, for a generic orientation of the molecule and  value of the applied field, we expect a larger mixing with $\ket{T_+^L}$ (maximized close to the avoided level crossing) and smaller with $\ket{T_-^L}$, as reported in Ref. \cite{dmi_chiesa_2026}. 
Note that a similar mechanism could also arise from the model proposed in Ref. \cite{Fay2021b}, where the traveling electron is also subject to isotropic  %FM/AFM 
exchange and DMI (although not mediated by vibrations). 

In general, the dynamics will be non-trivial, encompassing multi-frequency damped oscillations, with the relative weight of the different components depending on the microscopic Hamiltonian parameters and also on the CT rate. 
Therefore, to shed light on potential spectroscopic features of the $\ket{T_+^L}-\ket{T_-^L}$ population imbalance without limiting to a specific model system, we consider symmetry directions with the molecule parallel ($\theta=0, \pi$), or perpendicular ($\theta=\pi/2$) to the applied field. 
In the former case, $\mathbf{D} \parallel z$ and the CISS state has the same form $\ket{\psi_\parallel^\chi}$ of Eq. \eqref{eq:hore-state}. In the latter $D_z=0$ and hence we can consider a generic state of the form
\begin{eqnarray}
    &&\ket{\psi^L_\perp} = \cos \frac{\chi}{2} \ket{S} \\ \nonumber &+& \sin \frac{\chi}{2}    
      \left( \sqrt{\frac{1-w}{2}}  \ket{T_-^L}- \sqrt{\frac{1+w}{2}} \ket{T_+^L} \right) ,
\end{eqnarray}
%\begin{eqnarray}
%    \ket{\psi_\chi} &=& \cos \frac{\chi}{2} \ket{S} + \sin \frac{\chi}{2} \Big[ \cos \gamma  \;\ket{T_0} \\ \nonumber 
%    &+& \sin \gamma \left( \sqrt{\frac{1-w}{2}} e^{i\phi}  \ket{T_-}- \sqrt{\frac{1+w}{2}} e^{-i\phi} \ket{T_+} \right)\Big] ,
%\end{eqnarray}
where the parameter $w$ quantifies the relative weight of $\ket{T_+}$ vs $\ket{T_-}$ components, $\chi$ the CISS efficiency and we have considered for simplicity real coherences \footnote{this does not affect EPR simulations in the linear response regime since $\ket{T_\pm}$ are practically eigenstates}.
Since the value of $w$ depends on the microscopic parameters of the model (such as the spin-spin interactions, Zeeman gaps and CT time), we adopt this generic parametrization to make our conclusions more general. 
Due to the dependence on many parameters, realistic values for $w$ can range from $0-0.1$ (typically in small/large field limits compared to the exchange interactions) to values close to unity in proximity of avoided level crossings. In the following we vary $w$ in the realistic range from 0 to 0.7, which yields remarkable differences from the unpolarized $w=0$ case along specific directions. 
Finally, we keep the simulations for different $\theta$ separated to remove an additional $\theta-$dependent parameter from the state. 

Simulations at Q band for the benchmark system studied in the previous section are reported in Fig. \ref{fig:TpTmtoy}-(a), for different values of $w$ at $\theta=\pi/2$ and for $\theta=0,\pi$ in the bottom curves. Evidence of CISS appears for $\theta=\pi/2$, while at $\theta=0$ it only yields a renormalization of the intensity \cite{Eckvahl2023,Chiesa2025}.
By changing $w$ from 0 to 0.7 the larger population of the excited states  cancel the absorption peak in the perfect (100\%) CISS case, leaving only emission. 
It is worth noting that in isotropic solution this effect could be masked by spherical average. 
Indeed, the maximum at $\theta=0$ partially compensates the removed absorbance at $\theta=\pi/2$. Moreover, reduction of the CISS efficiency from 65 to 50\% (b vs a panel) also makes the effect less clear.

\begin{figure}
    \centering
    \includegraphics[width=0.9\linewidth]{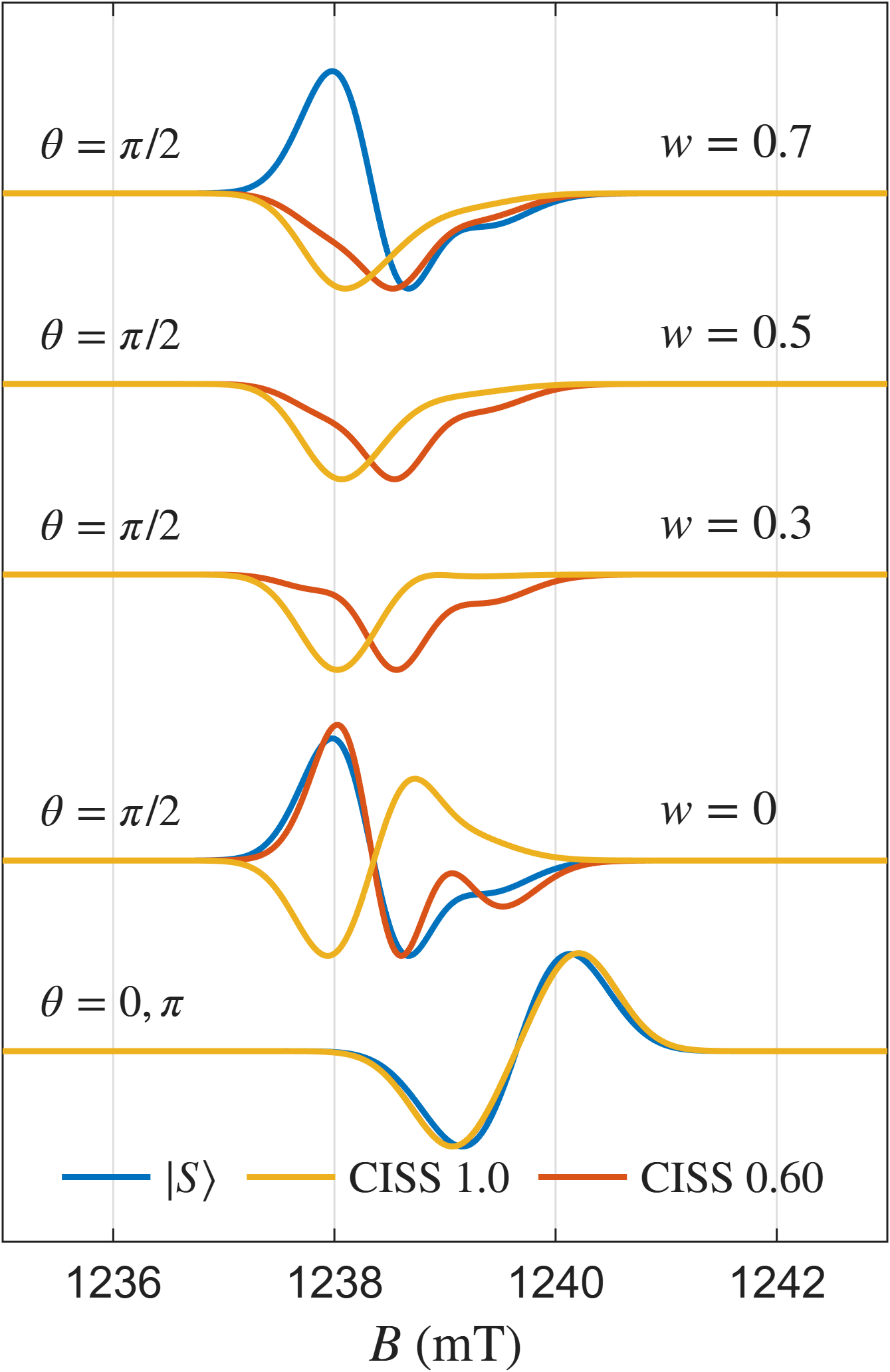}
    \caption{Simulations of time-resolved EPR spectra for the NDI-1G-Sd system of Ref. \cite{Latawiec2025} in Q band for different values of $w$ at $\theta=\pi/2$. Bottom lines are computed a reference for $\theta=0,\pi$.  Parameters used in the simulation are from \cite{Latawiec2025}.}
    \label{fig:TpTmHair}
\end{figure}

The situation reported in Fig. \ref{fig:TpTmtoy} is representative of experimental studies performed in Refs. \cite{Eckvahl2024,Latawiec2025}. For instance, Q-band simulations for the DNA hairpins reported in Ref. \cite{Latawiec2025} are shown in Fig. \ref{fig:TpTmHair} for different values of $w$. Similarly to the previous example, the absorption peak for the 100\% CISS simulation (blue curves) disappears in presence of a significant population imbalance between $\ket{T_+^L}$ and $\ket{T_-^L}$ (bottom vs top panel). However, the effect is partially masked by the spherical average, due to the maximum appearing for $\theta$ close to 0 (bottom lines in each panel). Moreover, reduction of CISS efficiency further reduces the visibility of the effect.

Our conclusion is that experimentally observed behaviours (explained so far with $w=0$) can be compatible with significant non-zero values of $w$, possibly by adjusting the value of the CISS efficiency.

Careful choice of the system or studies on samples aligned with LC could highlight the potential presence of a net spin polarization. This should also manifest as a non-zero integral of the spectrum.

%Comment on magnetic field dependence? According to \cite{dmi_chiesa_2026}, a strong magnetic field dependence may be accompanied by a large value of $w$. Otherwise large CISS efficiency in high field requires large DMI vector compared to J and Zeeman gap.

\section{Discussion and conclusions}

We have investigated spectral signatures of a broad class of radical pair spin states related to microscopic models for CISS in charge transfer.

In most CISS models (in small magnetic field) the only anisotropic term in the Hamiltonian is represented by the spin-orbit coupling, which thus typically determines the orientation of the predicted spin polarization \cite{Zhang2025}. This is also the case of models \cite{Fay2021b,dmi_chiesa_2026} in which the polarization arises from effective anti-symmetric spin-spin interactions between the unpaired spin on the donor and that of the transferred electron. The anisotropy of this interaction also  depends on the spin-orbit coupling vector. 
Ab-initio calculations on typical molecules displaying CISS indicate that the spin-orbit coupling may be not parallel to the charge transfer axis, potentially with a sizeable orthogonal component.
A similar observation results from atomistic models where the geometric spin--orbit is explicitly accounted for Ref.~\citenum{zhou2026real}, that predict sizeable transient spin polarization orthogonal to the CT axis.

In addition, recent models for CISS in ET \cite{dmi_chiesa_2026,Fay2021b} and experimental evidence of an important magnetic field dependence of the CISS efficiency in some cases \cite{Latawiec2025} indicate that the magnetic field applied to perform EPR measurements may be comparable to the microscopic spin-spin interactions driving spin polarization. In particular, by splitting the triplet sublevels, it could induce an $M-$dependent mixing of the different triplet sublevels (as noted in \cite{dmi_chiesa_2026}) with the consequent rise of a net spin-polarization. 
The mechanism proposed in Ref.~\citenum{Zhang2025}  relies instead on the interplay between multiple-paths hopping and spin-orbit coupling and strong spin decoherence acting solely on the transferred electron. It is therefore a single-electron model which only considers the dynamics of the transferred electron.  The two-spin density matrix obtained by starting from a singlet state could also in this case show a net spin polarization.

The effects of both a generically-oriented spin polarization (not parallel to the dipolar axis) and of a net spin polarization are hardly visible in measurements on randomly oriented solutions. Indeed, in these conditions they can be masked by the overlap between transitions from differently oriented molecules and would typically require very large CISS efficiency to become evident. Conversely, such spectral features would be more clearly distinguishable if molecules are oriented with liquid crystals. 
Measurements reported in Refs. \cite{Eckvahl2024,Latawiec2025,Latawiec2026} are on randomly oriented samples. These suggest a larger polarization along the charge-transfer axis, but cannot exclude a transverse component and neither a population imbalance between $\ket{T_+^L}$ and $\ket{T_-^L}$. 
So far, molecules oriented with LC were reported in \cite{Eckvahl2023}, where however CISS features are less evident. The absence of any difference between chiral and achiral molecules  in the parallel orientation indicates again a CISS polarization along the dipolar axis, but further studies on other systems oriented with LC are required to validate recent theoretical models. 
A recent study on achiral donor-acceptor molecules in cholesteric liquid crystal \cite{Eckvahl2026} also shows clear evidence of CISS with the magnetic field perpendicular to the dipolar axis and no evidence in the parallel orientation, consistently with weak or absent transverse polarization.

As a general conclusion, we can summarize which properties of the radical pair spin state are visible (by a different extent) by EPR. The clearest signature is usually (i) the triplet population. A net magnetization, associated to an imbalance between $\ket{T_+^L}$ and  $\ket{T_-^L}$ (ii), should yield a non-zero integral of the spectrum. This effect could be masked by an overlapping non-trivial background. Finally, real (iii) and imaginary (iv) singlet-triplet coherences can be evidenced by measurements in perfectly aligned samples and/or at short times, before losing the electronic spin coherences.

%Also in \cite{Eckvahl2026} (achiral molecules in chiral solvent) we do not see CISS features with field parallel to the dipolar axis, indicating weak or absent transverse polarization.
%, while the small asymmetry between the two lateral wings could be compatible with a small non-zero $\ket{T_+}-\ket{T_-}$ imbalance in the perpendicular arrangement.

%\section*{Methodology}
%All TREPR spectra were simulated using EasySpin (version 6.0.0-dev.53)\cite{easyspin} in MATLAB, via the \textit{pepper} function.

\section*{Acknowledgments}
\noindent 
The work was funded by the Horizon Europe Programme within the ERC-Synergy project CASTLE (proj. n. 101071533). Views and
opinions expressed are however those of the author(s) only and do not necessarily reflect those of the European Union or the European Commission. Neither the European Union nor
the granting authority can be held responsible for them. 
A.Ch. also acknowledges funding from the University of
Parma through the action Bando di Ateneo 2024 per la ricerca.
This research benefits from the High-Performance Computing
facility of the University of Parma, Italy. 

% save the normal section style (restored for the Supplementary Information)
\makeatletter
\let\SIsaved@hangfrom@section\@hangfrom@section
\let\SIsaved@sectioncntformat\@sectioncntformat
\makeatother
\appendix
\section{Symmetry Analysis of the Electronic DMI Contribution}
\label{app:localization}
\subsection{Calculation of $t$ and $\lambda$ for (R)-\textbf{1}}
First, a geometry optimization for (R)-\textbf{1} at the B3LYP/6-31G(d) level was performed using the Gaussian 16 package.~\cite{g16}
The hopping term $t$ was obtained by performing DFT calculations on the (R)-\textbf{1} molecule at the same level of theory and exploiting the Frontier Molecular Orbital (FMO) approach~\cite{FMO1,FMO2} within the QChem software, version 6.1 \cite{qchem}, to obtain the electronic coupling between the highest occupied orbital localized on the chiral bridge and the lowest unoccupied on the NDI acceptor. Finally, the spin-orbit coupling parameter $\lambda$ was determined from DFT calculations on the same orbital pair using the 2.12.1 PyScf package~\cite{pyscf} at the B3LYP/def2-TZVP level.

\subsection{DMI in achiral systems}
\label{app:symmetry}
As a high-symmetry achiral reference, consider system \textbf{2}  (Fig.~\ref{fig:scheme_coupling_jacs}a), which belongs to the $C_{2v}$ point group, with the $z$ axis along the CT direction, coinciding with the$C_2$ axis, and the $xz$/$yz$ planes as the two vertical mirror planes (Fig.~\ref{fig:c2v_matrices}a). Because the spin-independent electronic Hamiltonian is totally symmetric, its off-diagonal matrix elements can be nonzero only between orbitals of the same irreducible representation. The three components of the SOC pseudovector, by contrast, transform as the components of an axial vector: with the axis convention adopted here, $\mathrm{SOC}_x$, $\mathrm{SOC}_y$, and $\mathrm{SOC}_z$ transform as $B_2$, $B_1$, and $A_2$ respectively -- none of them totally symmetric ($A_1$). An orbital pair coupled by the spin-independent Hamiltonian can therefore never simultaneously be coupled by any SOC component, and the electronic DMI contribution vanishes for this system within the mechanism considered here. The symmetry irreps of the hopping and SOC operators, together with the corresponding selection rules, are summarized in Table~\ref{tab:c2v_selectionrules}.

\begin{table}[h]
\centering
\setlength{\tabcolsep}{10pt}
\begin{tabular}{lll}
\toprule
Operator & Irrep & Allowed Couplings \\
\midrule
Hopping & A$_1$ & States with same irrep \\
$\hat{H}_{SOC,x}$ &  B$_2$ & A$_1 \leftrightarrow$ B$_2$, A$_2 \leftrightarrow$ B$_1$\\
$\hat{H}_{SOC,y}$ &  B$_1$ & A$_1 \leftrightarrow$ B$_1$, A$_2 \leftrightarrow$ B$_2$\\
$\hat{H}_{SOC,z}$ & A$_2$ & A$_1 \leftrightarrow$ A$_2$, B$_2 \leftrightarrow$ B$_1$ \\
\bottomrule
\end{tabular}
\caption{Symmetry of hopping and SOC operators for the $C_{2v}$ point group and corresponding allowed state couplings.}
\label{tab:c2v_selectionrules}
\end{table}

To confirm the above symmetry arguments, we calculated the hopping and SOC
matrix elements between localized orbitals obtained from DFT on molecule \textbf{2}.
We first obtained the MOs of molecule \textbf{2} using the PyScf package at the B3LYP/def2-TZVP level, assigned an irrep to each MO, and divided them into irrep blocks $A_1$, $A_2$, $B_1$, and $B_2$. Since canonical MOs diagonalize the Kohn--Sham Hamiltonian, the hopping matrix elements vanish by construction in the MO basis; computing them therefore requires a localization procedure. A localized orbital basis was obtained by localizing the MOs onto the donor, bridge, and acceptor fragments with a Pipek--Mezey scheme applied independently within each irrep block, so as to preserve the symmetry character of the resulting orbitals. SOC matrix elements were computed on the same basis. 
The \textit{ab initio} calculations numerically confirm the symmetry predictions.
Fig.~\ref{fig:c2v_matrices}b shows the off-diagonal elements of the spin-independent Hamiltonian, with orbitals grouped by their $C_{2v}$ irreducible representation: as expected for a totally symmetric operator, nonzero matrix elements occur only within the diagonal irrep blocks. Figs.~\ref{fig:c2v_matrices}c--e show the complementary structure for the three SOC matrices, with couplings allowed only among certain irrep blocks as highlighted by Table~\ref{tab:c2v_selectionrules}. Consequently, no orbital pair exhibits simultaneously nonzero spin-independent hopping and SOC matrix elements, providing full numerical validation of the symmetry argument above and confirming a vanishing electronic DMI contribution for this $C_{2v}$ system within the present model.

The cancellation found for the $C_{2v}$ system is not a general
consequence of achirality. Consider instead an achiral molecule of $C_s$
symmetry, with the CT axis lying in the molecular mirror plane. Orbitals
in $C_s$ symmetry belong to either the $A'$ or $A''$ irreducible
representation -- symmetric or antisymmetric, respectively, under
reflection through the plane. The component of an axial vector normal to
the mirror plane transforms as $A'$, while its in-plane components
transform as $A''$. The normal component of the SOC pseudovector can
therefore couple orbitals of the same irrep, and may be nonzero for a pair
also coupled by the spin-independent Hamiltonian: an electronic DMI
contribution is symmetry-allowed here, oriented normal to the mirror plane
and hence perpendicular to the CT axis.

Thus, within the present model, molecular achirality alone is not sufficient to exclude a transverse spin polarization. This does not imply that a sizeable or experimentally detectable polarization must arise in
any given $C_s$ molecule -- only that such a contribution is not symmetry-forbidden; additional dynamical effects or conformational averaging may strongly suppress it in practice.
%apsrev4-2.bst 2019-01-14 (MD) hand-edited version of apsrev4-1.bst
%Control: key (0)
%Control: author (8) initials jnrlst
%Control: editor formatted (1) identically to author
%Control: production of article title (0) allowed
%Control: page (0) single
%Control: year (1) truncated
%Control: production of eprint (0) enabled
%

%%%%%%%%%%%%%%%%%%%%%%%%%%%%%%%%%%%%%%%%%%%%%%%%%%%%%%%%%%%%%%%%%%%%%%%%%%%%%%
%%%%%%%%%%%%%%%%%%%%%%%%%%   SUPPLEMENTARY INFORMATION   %%%%%%%%%%%%%%%%%%%%%%%%
%%%%%%%%%%%%%%%%%%%%%%%%%%%%%%%%%%%%%%%%%%%%%%%%%%%%%%%%%%%%%%%%%%%%%%%%%%%%%%

\clearpage
\onecolumngrid

% restore the standard (non-appendix) section style and use S-numbering
\makeatletter
\let\@hangfrom@section\SIsaved@hangfrom@section
\let\@sectioncntformat\SIsaved@sectioncntformat
\makeatother
\setcounter{section}{0}
\setcounter{figure}{0}
\setcounter{table}{0}
\setcounter{equation}{0}
\renewcommand{\thesection}{S\arabic{section}}
\renewcommand{\thefigure}{S\arabic{figure}}
\renewcommand{\thetable}{S\arabic{table}}
\renewcommand{\theequation}{S\arabic{equation}}
\renewcommand{\theHsection}{S\arabic{section}}
\renewcommand{\theHfigure}{S\arabic{figure}}
\renewcommand{\theHtable}{S\arabic{table}}
\renewcommand{\theHequation}{S\arabic{equation}}

\begin{center}
  {\large\bfseries Supporting Information}
\end{center}

\section{TREPR simulations on the (R)-\textbf{1} and \textbf{2} molecules}
We simulated the TREPR in X-band for the (R)-\textbf{1} and \textbf{2} molecules reported in Ref.~\citenum{Eckvahl2024}.
The parameters used to simulate both chiral and achiral versions are reported in Ref.~\citenum{Eckvahl2024}. The hyperfine interaction was not included for simplicity and we used a linewidth peak-to-peak of 0.2 mT.

\begin{figure*}[ht]
    \centering
    \includegraphics[width=\linewidth]{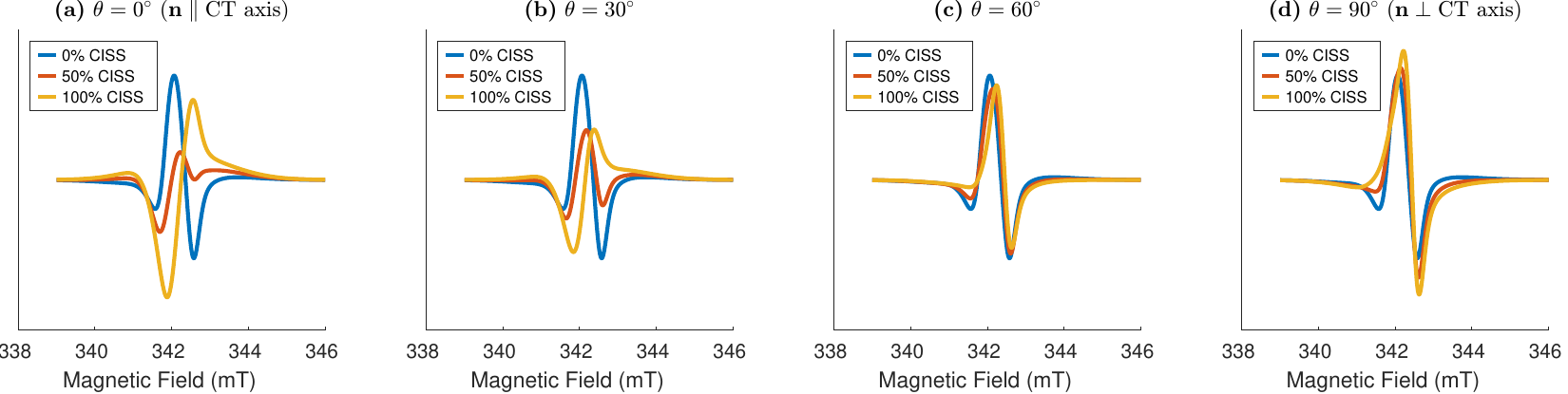}
    \caption{\textbf{Simulated powder TREPR spectra for an initial triplet polarization along different molecular-frame directions for molecules (R)-\textbf{1} and \textbf{2} in X-band}. The directions are parametrized by the polar angle $\alpha$ from the CT axis ($\beta = 0$ throughout). Panel (a): polarization direction parallel to the CT axis ($\alpha=0^\circ$); panel (d): polarization direction orthogonal to the CT axis ($\alpha=90^\circ$); panels (b,c) show intermediate orientations. Each panel reports spectra for CISS contributions of 0\%, 50\%, and 100\%. The spectra are normalized to the minimum of each signal.}
    \label{fig:jacs_powder_direction}
\end{figure*}

\begin{figure*}[ht!]
	\centering
	\includegraphics[width=0.7\linewidth]{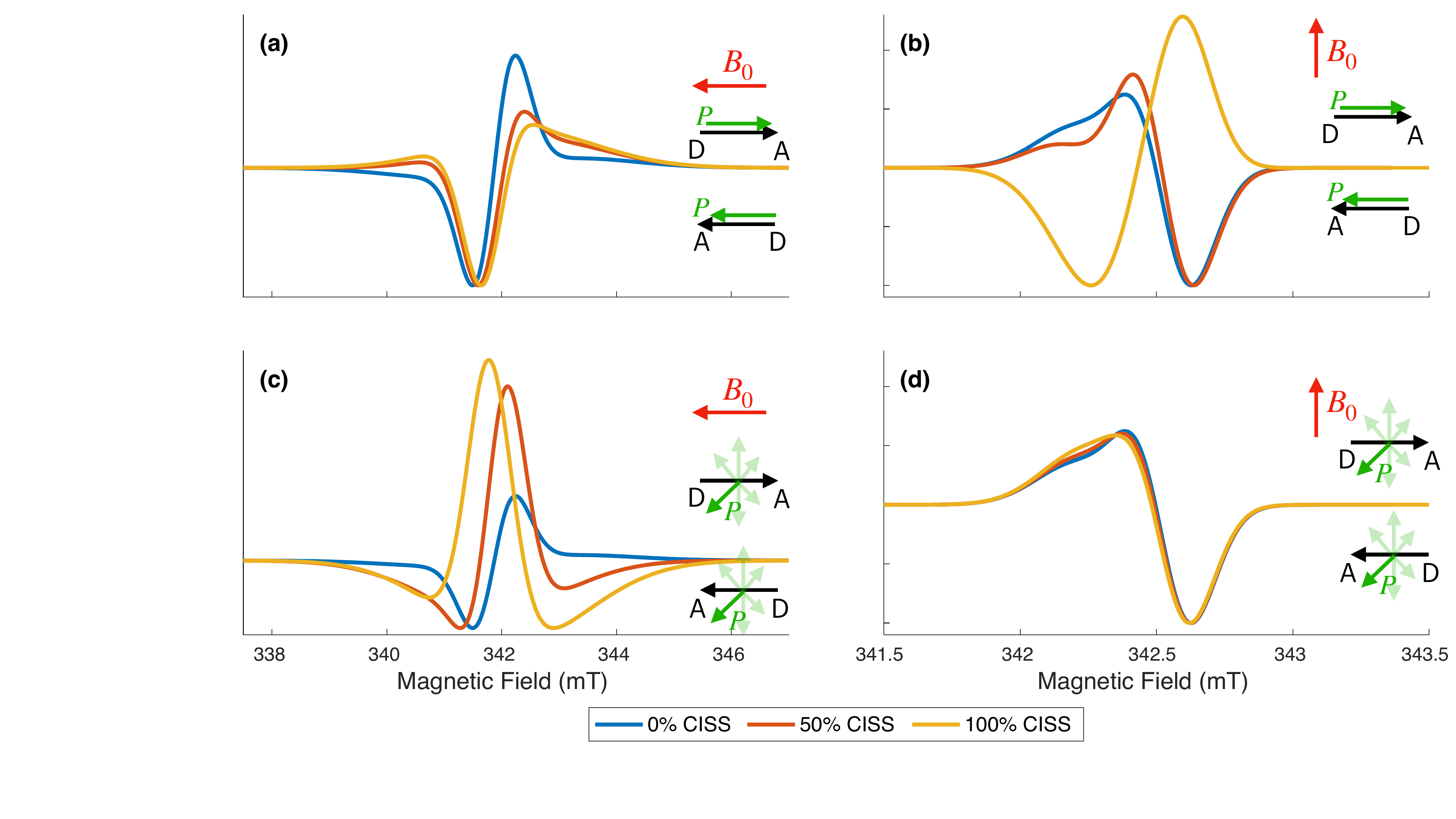}
	\caption{\textbf{Simulated oriented-sample TREPR spectra for the four combinations of polarization orientation relative to the dipolar axis, for molecules (R)-\textbf{1} and \textbf{2}, in X-band.} Rows: orientation of the polarization vector relative to the molecular dipolar axis (parallel, top; perpendicular, bottom). Columns: orientation of the external magnetic field $\mathbf{B}_0$ relative to the dipolar axis ($\mathbf{B}_0 \parallel$, left; $\mathbf{B}_0 \perp$, right). In each panel, spectra are shown for CISS contributions of 0\%, 50\%, and 100\%. The spectra are normalized to the minimum of each signal.}
	\label{fig:jacs_epr_oriented}
\end{figure*}

\clearpage
\newpage

\section{\textit{Ab initio} validation of the symmetry-based selection rules}
\begin{figure}[ht!]
	\centering
	\includegraphics[width=0.9\linewidth]{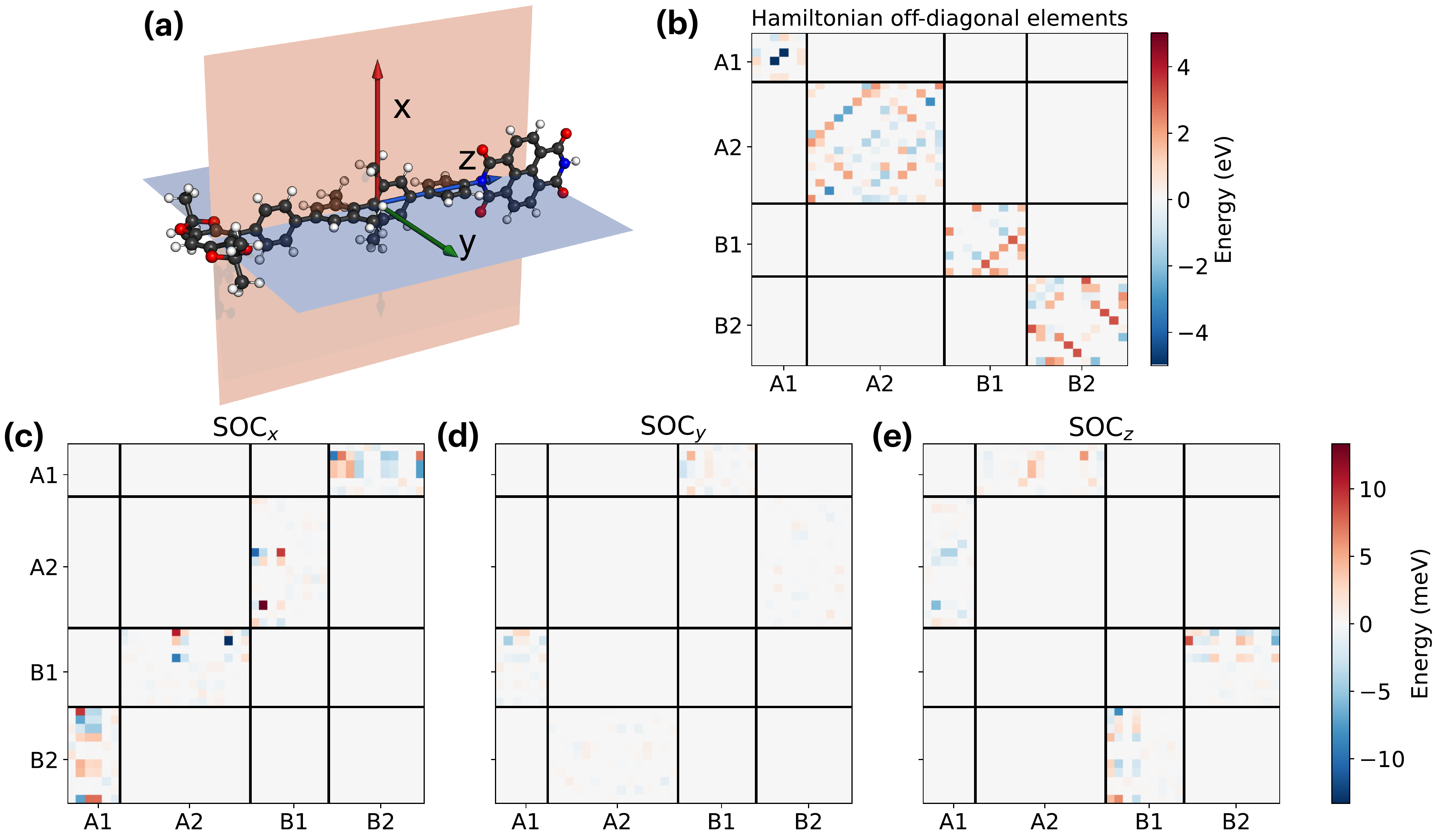}
	\caption{\textbf{The C$_{2v}$ molecule, its symmetry planes, and heat maps of the Hamiltonian and spin–orbit coupling matrix elements resolved by irreducible representations.} \textbf{a:} Molecular structure of the achiral system introduced in Ref.~\cite{Eckvahl2024}. The charge-transfer (CT) axis is aligned along the $z$ direction, while the two vertical mirror planes, $\sigma_v(xz)$ (red) and $\sigma_v(yz)$ (blue), are highlighted. \textbf{b:} Off-diagonal matrix elements of the Kohn–Sham Hamiltonian in the localized orbital basis, grouped according to the irreducible representations of the C$_{2v}$ point group. \textbf{c–e:} Matrix elements of the one-electron spin–orbit coupling operator, $\mathrm{SOC}_x$, $\mathrm{SOC}_y$, and $\mathrm{SOC}_z$, respectively, represented in the same localized basis. To improve readability, the localization procedure was restricted to the HOMO $\pm$ 20 molecular orbitals.}
	\label{fig:c2v_matrices}
\end{figure}

\clearpage
\newpage

\section{Input geometries}\label{sec:inputgeometry_SI}
Molecule (R)-\textbf{1}
\renewcommand{\arraystretch}{0.6}
\begin{longtable}{ccrrr}
    \hline
    \textbf{Atom No.} & \textbf{Element} & \textbf{x} & \textbf{y} & \textbf{z} \\
    \hline
1 & C & 5.402122 & 0.655564 & -16.642124 \\
2 & C & 4.289526 & 0.988009 & -17.407433 \\
3 & C & 3.157751 & 1.274395 & -16.651695 \\
4 & C & 3.137616 & 1.227892 & -15.261780 \\
5 & C & 4.252552 & 0.899362 & -14.471792 \\
6 & C & 5.387155 & 0.617896 & -15.251902 \\
7 & O & 6.640975 & 0.274106 & -14.792039 \\
8 & C & 7.414360 & -0.107518 & -15.958460 \\
9 & O & 6.655545 & 0.339716 & -17.109552 \\
10 & O & 1.916258 & 1.619225 & -17.130483 \\
11 & C & 1.130240 & 2.000384 & -15.973885 \\
12 & O & 1.872244 & 1.544247 & -14.813887 \\
13 & C & 8.740286 & 0.627198 & -15.926343 \\
14 & C & 7.527087 & -1.625682 & -15.997755 \\
15 & C & -0.199027 & 1.272557 & -16.019703 \\
16 & C & 1.024736 & 3.518766 & -15.923728 \\
17 & C & 4.234221 & 0.855424 & -12.994061 \\
18 & C & 5.391814 & 1.137741 & -12.246370 \\
19 & C & 5.370448 & 1.096676 & -10.854966 \\
20 & C & 4.199474 & 0.770924 & -10.157406 \\
21 & C & 3.046715 & 0.489124 & -10.901226 \\
22 & C & 3.059256 & 0.530557 & -12.293094 \\
23 & C & 4.175196 & 0.728724 & -8.662582 \\
24 & C & 4.381688 & -0.493996 & -7.992899 \\
25 & C & 4.327897 & -0.535461 & -6.587121 \\
26 & C & 4.069790 & 0.649640 & -5.867258 \\
27 & C & 3.895002 & 1.857279 & -6.554993 \\
28 & C & 3.939727 & 1.919419 & -7.954769 \\
29 & C & 4.003476 & 0.657205 & -4.372831 \\
30 & C & 2.874023 & 0.168597 & -3.683637 \\
31 & C & 2.855115 & 0.195676 & -2.276515 \\
32 & C & 3.965248 & 0.709081 & -1.577096 \\
33 & C & 5.086864 & 1.219235 & -2.251726 \\
34 & C & 5.082624 & 1.186851 & -3.653388 \\
35 & O & 3.727993 & 3.028948 & -5.837781 \\
36 & O & 6.148299 & 1.743726 & -4.336731 \\
37 & C & 4.635541 & -1.751195 & -8.814332 \\
38 & C & 5.208844 & -2.916869 & -8.002983 \\
39 & C & 4.429338 & -3.084321 & -6.698661 \\
40 & C & 4.592907 & -1.832503 & -5.832912 \\
41 & C & 1.671401 & -0.336347 & -4.470955 \\
42 & C & 0.726863 & -1.213041 & -3.644754 \\
43 & C & 0.396963 & -0.517202 & -2.324254 \\
44 & C & 1.668157 & -0.342745 & -1.488116 \\
45 & C & 3.738361 & 3.231836 & -8.675871 \\
46 & C & 6.265756 & 1.786990 & -1.496046 \\
47 & C & 7.158033 & 0.823024 & -4.739684 \\
48 & C & 2.386355 & 3.319423 & -5.458411 \\
49 & H & 4.303254 & 1.021027 & -18.489287 \\
50 & H & 8.557986 & 1.704434 & -15.898267 \\
51 & H & 9.324063 & 0.385062 & -16.819162 \\
52 & H & 9.308112 & 0.335055 & -15.038139 \\
53 & H & 6.524215 & -2.062133 & -16.014813 \\
54 & H & 8.056391 & -1.986513 & -15.110996 \\
55 & H & 8.067647 & -1.939817 & -16.895510 \\
56 & H & -0.021946 & 0.194576 & -16.052383 \\
57 & H & -0.758908 & 1.571543 & -16.910593 \\
58 & H & -0.788047 & 1.513806 & -15.130022 \\
59 & H & 2.029446 & 3.949874 & -15.888769 \\
60 & H & 0.474443 & 3.828166 & -15.030345 \\
61 & H & 0.508663 & 3.889801 & -16.814151 \\
62 & H & 6.312141 & 1.391893 & -12.758932 \\
63 & H & 6.274748 & 1.322241 & -10.295180 \\
64 & H & 2.128825 & 0.231573 & -10.378615 \\
65 & H & 2.151746 & 0.308372 & -12.842116 \\
66 & H & 5.297390 & -1.515077 & -9.654596 \\
67 & H & 3.687016 & -2.066005 & -9.275087 \\
68 & H & 6.264777 & -2.720810 & -7.766517 \\
69 & H & 5.183017 & -3.834021 & -8.603754 \\
70 & H & 4.768817 & -3.966773 & -6.142719 \\
71 & H & 3.365081 & -3.240440 & -6.927788 \\
72 & H & 5.620066 & -1.806970 & -5.437700 \\
73 & H & 3.951700 & -1.897299 & -4.949227 \\
74 & H & 2.000111 & -0.872287 & -5.365740 \\
75 & H & 1.109254 & 0.531060 & -4.849505 \\
76 & H & 1.205590 & -2.181136 & -3.437002 \\
77 & H & -0.182182 & -1.423950 & -4.220970 \\
78 & H & -0.348103 & -1.081297 & -1.750556 \\
79 & H & -0.044190 & 0.467074 & -2.538098 \\
80 & H & 1.947950 & -1.314453 & -1.053482 \\
81 & H & 1.466827 & 0.307455 & -0.629854 \\
82 & H & 3.697550 & 4.059710 & -7.966448 \\
83 & H & 2.811284 & 3.222231 & -9.262484 \\
84 & H & 4.551006 & 3.421602 & -9.384885 \\
85 & H & 5.945324 & 2.544010 & -0.772379 \\
86 & H & 6.978131 & 2.246227 & -2.182829 \\
87 & H & 6.781466 & 1.007739 & -0.921211 \\
88 & H & 7.974039 & 1.421818 & -5.152408 \\
89 & H & 6.788802 & 0.139469 & -5.513495 \\
90 & H & 7.531555 & 0.239111 & -3.886427 \\
91 & H & 2.017171 & 2.601465 & -4.716086 \\
92 & H & 1.714316 & 3.319980 & -6.328405 \\
93 & H & 2.400707 & 4.317497 & -5.012752 \\
94 & C & 3.963207 & 0.711634 & -0.081055 \\
95 & C & 3.533333 & 1.837207 & 0.635736 \\
96 & C & 3.536598 & 1.841546 & 2.028591 \\
97 & C & 3.967412 & 0.709880 & 2.717808 \\
98 & C & 4.397182 & -0.419985 & 2.025976 \\
99 & C & 4.396593 & -0.413474 & 0.632710 \\
100 & H & 3.193647 & 2.714448 & 0.092291 \\
101 & H & 3.198158 & 2.711530 & 2.581877 \\
102 & H & 4.737525 & -1.290629 & 2.577069 \\
103 & H & 4.734990 & -1.290222 & 0.087711 \\
104 & N & 3.967662 & 0.708811 & 4.163821 \\
105 & C & 2.712559 & 0.711307 & 4.811419 \\
106 & C & 2.734542 & 0.710624 & 6.298876 \\
107 & C & 3.967467 & 0.710467 & 6.993955 \\
108 & C & 5.200779 & 0.708655 & 6.299189 \\
109 & C & 5.222899 & 0.704809 & 4.811744 \\
110 & C & 1.540187 & 0.710428 & 7.001778 \\
111 & C & 1.540308 & 0.710807 & 8.409396 \\
112 & C & 2.735018 & 0.712175 & 9.111719 \\
113 & C & 3.967284 & 0.712100 & 8.414962 \\
114 & C & 5.200052 & 0.713636 & 9.110167 \\
115 & C & 6.395309 & 0.713360 & 8.409448 \\
116 & C & 6.395435 & 0.710428 & 7.001778 \\
117 & C & 2.715355 & 0.713757 & 10.599955 \\
118 & N & 3.965714 & 0.715729 & 11.241614 \\
119 & C & 5.210972 & 0.715486 & 10.596108 \\
120 & C & 4.005526 & 0.717763 & 12.709716 \\
121 & O & 6.244348 & 0.716792 & 11.248556 \\
122 & O & 1.670924 & 0.713573 & 11.234033 \\
123 & O & 6.261903 & 0.696851 & 4.174826 \\
124 & O & 1.673408 & 0.715895 & 4.174475 \\
125 & H & 0.609733 & 0.710335 & 6.444192 \\
126 & H & 0.609729 & 0.710650 & 8.966708 \\
127 & H & 7.325331 & 0.714940 & 8.967901 \\
128 & H & 7.325689 & 0.709396 & 6.443857 \\
129 & H & 2.981150 & 0.719429 & 13.074393 \\
130 & H & 4.540892 & 1.603363 & 13.059712 \\
131 & H & 4.538805 & -0.168063 & 13.062407 \\
    \caption{Atomic coordinates of the chiral ((R) enantiomer) molecule in \cite{Eckvahl2024} in Angstrom.}
\end{longtable}

Molecule \textbf{2}
\renewcommand{\arraystretch}{0.6}
\begin{longtable}{ccrrr}
    \hline
    \textbf{Atom No.} & \textbf{Element} & \textbf{x} & \textbf{y} & \textbf{z} \\
    \hline
1 & C & -1.17283500 & 0.00000000 & 12.62213009 \\
2 & C & 0.00000000 & 0.00000000 & 13.38636758 \\
3 & C & 1.17283500 & 0.00000000 & 12.62213009 \\
4 & C & 1.15782121 & -0.00000000 & 11.23797881 \\
5 & C & 0.00000000 & 0.00000000 & 10.44562152 \\
6 & C & -1.15782121 & -0.00000000 & 11.23797881 \\
7 & O & -2.47328514 & 0.00000000 & 10.77810041 \\
8 & C & -3.33436790 & 0.00000000 & 11.94424613 \\
9 & O & -2.46034969 & 0.00000000 & 13.09793729 \\
10 & O & 2.46034969 & 0.00000000 & 13.09793729 \\
11 & C & 3.33436790 & 0.00000000 & 11.94424613 \\
12 & O & 2.47328514 & 0.00000000 & 10.77810041 \\
13 & C & -4.25046935 & 1.25575915 & 11.95226093 \\
14 & C & -4.25046935 & -1.25575915 & 11.95226093 \\
15 & C & 4.25046935 & -1.25575915 & 11.95226093 \\
16 & C & 4.25046935 & 1.25575915 & 11.95226093 \\
17 & C & 0.00000000 & 0.00000000 & 8.96552215 \\
18 & C & 0.00000000 & 1.20226367 & 8.23402806 \\
19 & C & 0.00000000 & 1.19845139 & 6.84351612 \\
20 & C & 0.00000000 & 0.00000000 & 6.11204394 \\
21 & C & 0.00000000 & -1.19845139 & 6.84351612 \\
22 & C & 0.00000000 & -1.20226367 & 8.23402806 \\
23 & C & 0.00000000 & 0.00000000 & 4.62880701 \\
24 & C & 1.20528613 & -0.00000000 & 3.90552095 \\
25 & C & 1.22726425 & -0.00000000 & 2.50831698 \\
26 & C & 0.00000000 & 0.00000000 & 1.79720831 \\
27 & C & -1.22726425 & -0.00000000 & 2.50831698 \\
28 & C & -1.20528613 & -0.00000000 & 3.90552095 \\
29 & C & 0.00000000 & 0.00000000 & 0.29654863 \\
30 & C & 0.00000000 & 1.21348760 & -0.41498864 \\
31 & C & 0.00000000 & 1.19099934 & -1.81232456 \\
32 & C & 0.00000000 & 0.00000000 & -2.53291851 \\
33 & C & 0.00000000 & -1.19099934 & -1.81232456 \\
34 & C & 0.00000000 & -1.21348760 & -0.41498864 \\
35 & C & 0.00000000 & 0.00000000 & -4.01761520 \\
36 & C & 1.21442035 & -0.00000000 & -4.74294378 \\
37 & C & 1.21703134 & -0.00000000 & -6.13473121 \\
38 & C & 0.00000000 & 0.00000000 & -6.82932934 \\
39 & C & -1.21703134 & -0.00000000 & -6.13473121 \\
40 & C & -1.21442035 & -0.00000000 & -4.74294378 \\
41 & C & -2.53500695 & 0.00000000 & 1.77520855 \\
42 & C & 2.53500695 & 0.00000000 & 1.77520855 \\
43 & C & 0.00000000 & 2.53964566 & 0.31093001 \\
44 & C & 0.00000000 & -2.53964566 & 0.31093001 \\
45 & N & 0.00000000 & 0.00000000 & -8.27720797 \\
46 & C & 0.00000000 & 1.24596851 & -8.92327193 \\
47 & C & 0.00000000 & 1.22846063 & -10.41110379 \\
48 & C & 0.00000000 & 0.00000000 & -11.11029836 \\
49 & C & 0.00000000 & -1.22846063 & -10.41110379 \\
50 & C & 0.00000000 & -1.24596851 & -8.92327193 \\
51 & C & 0.00000000 & 2.42341134 & -11.11086239 \\
52 & C & 0.00000000 & 2.42772384 & -12.51830253 \\
53 & C & 0.00000000 & 1.23704400 & -13.22470407 \\
54 & C & 0.00000000 & 0.00000000 & -12.53157892 \\
55 & C & 0.00000000 & -1.23704400 & -13.22470407 \\
56 & C & 0.00000000 & -2.42772384 & -12.51830253 \\
57 & C & 0.00000000 & -2.42341134 & -11.11086239 \\
58 & C & 0.00000000 & 1.26096880 & -14.71225880 \\
59 & N & 0.00000000 & 0.00000000 & -15.35849307 \\
60 & C & 0.00000000 & -1.26096880 & -14.71225880 \\
61 & O & 0.00000000 & -2.26024348 & -15.37233056 \\
62 & O & -0.00000000 & 2.26024348 & -15.37233056 \\
63 & O & 0.00000000 & -2.30058522 & -8.29769563 \\
64 & O & -0.00000000 & 2.30058522 & -8.29769563 \\
65 & H & 0.00000000 & 0.00000000 & 14.46907547 \\
66 & H & -3.64945420 & 2.16912271 & 11.95016525 \\
67 & H & -4.87955917 & 1.24726935 & 12.84750228 \\
68 & H & -4.89367834 & 1.25044317 & 11.06703710 \\
69 & H & -3.64945420 & -2.16912271 & 11.95016525 \\
70 & H & -4.89367834 & -1.25044317 & 11.06703710 \\
71 & H & -4.87955917 & -1.24726935 & 12.84750228 \\
72 & H & 3.64945420 & -2.16912271 & 11.95016525 \\
73 & H & 4.87955917 & -1.24726935 & 12.84750228 \\
74 & H & 4.89367834 & -1.25044317 & 11.06703710 \\
75 & H & 3.64945420 & 2.16912271 & 11.95016525 \\
76 & H & 4.89367834 & 1.25044317 & 11.06703710 \\
77 & H & 4.87955917 & 1.24726935 & 12.84750228 \\
78 & H & 0.00000000 & 2.14794121 & 8.75941776 \\
79 & H & 0.00000000 & 2.14452037 & 6.31409356 \\
80 & H & 0.00000000 & -2.14452037 & 6.31409356 \\
81 & H & 0.00000000 & -2.14794121 & 8.75941776 \\
82 & H & 2.14956231 & -0.00000000 & 4.44381649 \\
83 & H & -2.14956231 & -0.00000000 & 4.44381649 \\
84 & H & 0.00000000 & 2.13203833 & -2.35096592 \\
85 & H & 0.00000000 & -2.13203833 & -2.35096592 \\
86 & H & 2.16058631 & -0.00000000 & -4.21066528 \\
87 & H & 2.15376212 & -0.00000000 & -6.68181948 \\
88 & H & -2.15376212 & -0.00000000 & -6.68181948 \\
89 & H & -2.16058631 & -0.00000000 & -4.21066528 \\
90 & H & -2.59394548 & 0.96162792 & 1.09434129 \\
91 & H & -3.36627898 & 0.00000000 & 2.48308051 \\
92 & H & -2.59394548 & -0.96162792 & 1.09434129 \\
93 & H & 3.36627898 & 0.00000000 & 2.48308051 \\
94 & H & 2.59394548 & 0.96162792 & 1.09434129 \\
95 & H & 2.59394548 & -0.96162792 & 1.09434129 \\
96 & H & -0.75701961 & 2.71019318 & 0.92586878 \\
97 & H & 0.00000000 & 3.36895605 & -0.39852328 \\
98 & H & 0.75701961 & 2.71019318 & 0.92586878 \\
99 & H & 0.00000000 & -3.36895605 & -0.39852328 \\
100 & H & -0.75701961 & -2.71019318 & 0.92586878 \\
101 & H & 0.75701961 & -2.71019318 & 0.92586878 \\
102 & H & 0.00000000 & 3.35162707 & -10.55168056 \\
103 & H & 0.00000000 & 3.35957782 & -13.07167909 \\
104 & H & 0.00000000 & -3.35957782 & -13.07167909 \\
105 & H & 0.00000000 & -3.35162707 & -10.55168056 \\
106 & H & 0.00000000 & 0.00000000 & -16.35830386 \\
    \caption{Atomic coordinates of the symmetrized achiral molecule in \cite{Eckvahl2024} in Angstrom.}
\end{longtable}

\end{document}